\documentclass[letterpaper,twocolumn,10pt]{article}
\usepackage{usenix,epsfig}
\usepackage{authblk}
\usepackage{todonotes}
\usepackage{tabularray}
\usepackage{listings}
\usepackage{amsmath}
\usepackage{subcaption}
\usepackage{enumitem}
\usepackage{titlesec}
\usepackage{booktabs}
\usepackage{multirow}
\usepackage{pdflscape}
\usepackage{placeins}
\usepackage{titlesec}
\usepackage[title]{appendix}

\definecolor{codegreen}{rgb}{0,0.6,0}
\definecolor{codegray}{rgb}{0.5,0.5,0.5}
\definecolor{codepurple}{rgb}{0.58,0,0.82}
\definecolor{backcolour}{rgb}{0.95,0.95,0.92}
\definecolor{pinkcolour}{HTML}{de90c0}

\lstdefinestyle{mystyle}{
    backgroundcolor=\color{backcolour},   
    commentstyle=\color{codegreen},
    keywordstyle=\color{magenta},
    numberstyle=\tiny\color{codegray},
    stringstyle=\color{codepurple},
    basicstyle=\ttfamily\footnotesize,
    breakatwhitespace=false,         
    breaklines=true,                 
    captionpos=b,                    
    keepspaces=true,                 
    numbers=left,                    
    numbersep=2pt,                  
    showspaces=false,                
    showstringspaces=false,
    showtabs=false,                  
    tabsize=1
}

\definecolor{pcmCodeBg}{HTML}{F5F7F9}
\definecolor{pcmCodeRule}{HTML}{D4DCE3}
\definecolor{pcmCodeKeyword}{HTML}{174A7E}
\definecolor{pcmCodeAPI}{HTML}{086B68}
\definecolor{pcmCodeObject}{HTML}{A14516}
\definecolor{pcmCodeConstant}{HTML}{6B3FA0}
\definecolor{pcmCodeComment}{HTML}{526C55}
\definecolor{pcmCodeString}{HTML}{7B3F80}
\lstdefinestyle{pcmcode}{
    language=C,
    basicstyle=\ttfamily\footnotesize,
    backgroundcolor=\color{pcmCodeBg},
    keywordstyle=\color{pcmCodeKeyword},
    commentstyle=\color{pcmCodeComment},
    stringstyle=\color{pcmCodeString},
    morekeywords={uint,int,float},
    emph={get_signal,get_signal_trigger_mask,get_control,set_control,__builtin_ctz,
          get_var,set_var,get_var_uint,set_var_uint,get_arr,set_arr,get_arr_uint,set_arr_uint,
          get_bitmap_idx, set_bitmap_idx,
          min, max},
    emphstyle=\color{pcmCodeAPI},
    emph=[2]{ACK,NACK,ACKED_BYTES,NACKED_BYTES,NUM_ACK,NUM_ECN,
             ACK_EV,RTT,CWND,
             PULL_TARGET,PULL_NO,NEEDS_MORE,HIGHEST_PULL_TARGET,LATEST_PULL,
             PREV_ACKED_BYTES,PREV_NACKED_BYTES,PREV_ECN,
             EVC,EVC_BITMAP,EVC_HEAD_IDX,EVC_NUM_VALID_EVS},
    emphstyle=[2]\color{pcmCodeObject},
    emph=[3]{MSS,MTU,EVC_SIZE,CREDITS_PER_PULL,signals_offset,trigger_mask_offset,
             DATA_ACK_BYTES,SUM,DELTA,CWND_BYTES,DATA_NACK_BYTES},
    emphstyle=[3]\color{pcmCodeConstant},
    numbers=left,
    numberstyle=\tiny\color{codegray},
    numbersep=6pt,
    stepnumber=1,
    firstnumber=1,
    numberblanklines=false,
    frame=single,
    rulecolor=\color{pcmCodeRule},
    framerule=0.3pt,
    framesep=3pt,
    framexleftmargin=14pt,
    xleftmargin=17pt,
    xrightmargin=3pt,
    columns=fullflexible,
    keepspaces=true,
    breaklines=true,
    breakatwhitespace=true,
    showstringspaces=false,
    tabsize=4,
    captionpos=b
}
\lstdefinelanguage{PCMYAML}{
    morekeywords={signals,variables,name,type,accumulation,
        trigger_type,trigger_op,trigger_threshold,
        reset_upon_trigger,initial_value,value,controls,control},
    morekeywords=[2]{uint,false},
    keywordstyle=[2]\color{pcmCodeAPI},
    morecomment=[l]{\#},
    morestring=[b]"
}

\titlespacing*{\section}
  {0pt}      
  {0.6em}    
  {0.3em}    

\titlespacing*{\subsection}
  {0pt}
  {0.5em}
  {0.3em}

\titlespacing*{\subsubsection}
  {0pt}
  {0.3em}
  {0.3em}

\titlespacing*{\paragraph}
  {0pt}      
  {0.5ex}    
  {0.5ex}    

\begin{document}

\date{}

\title{PCM: Programmable Congestion Management for Terabit RDMA NICs}

\author[1]{Mikhail Khalilov}
\author[1]{Timo Schneider}
\author[1]{Tommaso Bonato}
\author[1]{\authorcr Siyuan Shen}
\author[1]{Marcin Chrapek}
\author[1]{Yehor Kozyr}
\author[2]{Alberto Merino}
\author[3]{\authorcr Rip Sohan}
\author[3]{Surendra Anubolu}
\author[3]{Mohan Kalkunte}
\author[4]{\authorcr Ahmad Ghalayini}
\author[4]{Abdul Kabbani}
\author[1,4]{Torsten Hoefler}
\affil[1]{ETH Zürich}
\affil[2]{Universidad de Castilla-La Mancha}
\affil[3]{Broadcom Inc.}
\affil[4]{Microsoft Corporation}
{
    \makeatletter
    \renewcommand\AB@affilsepx{: \protect\Affilfont}
    \makeatother

    \affil[ ]{}

    \makeatletter
    \renewcommand\AB@affilsepx{, \protect\Affilfont}
    \makeatother
}

\maketitle

\thispagestyle{empty}

\subsection*{Abstract}

Datacenter congestion control (CC) and load balancing (LB) evolve rapidly as workload demands change, but deployment depends on RDMA hardware and transport interfaces. Emerging programmable RDMA engines tie policy code to vendor APIs. Operators must therefore port each policy for every platform they deploy. We address this with PCM, a programming model that separates portable congestion management from platform integration on datacenter NICs. Developers write event-driven handlers that they can debug in simulation and reuse on real hardware. PCM exposes flow telemetry as signals and uses triggers to support per-event reactions and coarser responses to accumulated feedback. Vendors provide optimized backends that map signals and output controls to native NIC mechanisms with defined feedback and control semantics. We demonstrate PCM's expressivity by implementing eleven well-known datacenter CC and LB algorithms. We deploy PCM across four distinct datapaths: an Ultra Ethernet simulator, CPU-based UCCL, and on-NIC offloads on NVIDIA ConnectX-8 and Microsoft 1P. On CX-8, PCM-Swift achieves processing capacity equivalent to one reaction per three 4~KiB packets at 800~Gbit/s, using one-fifth of the NIC's hardware threads. In UCCL, it matches native Swift's 1.56~Tbit/s goodput with 64~KiB chunk-level control. PCM's design informed ongoing Ultra Ethernet work toward a portable congestion management interface.

\section{Introduction}

\begin{figure}[t]
 \centering
 \includegraphics[width=\linewidth]{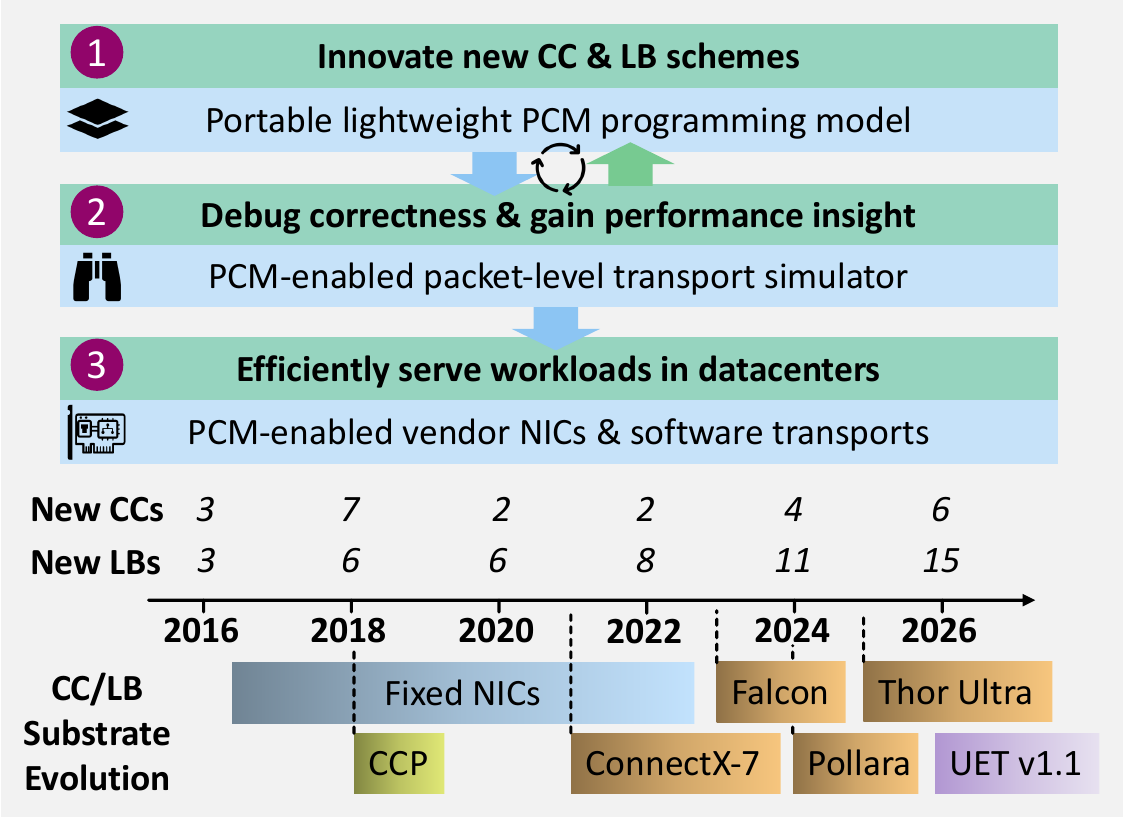}
 \caption{PCM generalizes decades of NIC congestion management evolution into a reusable workflow: tune policies in simulation and deploy them unchanged through vendor-optimized platform backends.}
 \label{fig:poster}
\end{figure}

Datacenter operators seek workload-specific congestion control and load balancing~\cite{dctcp,dcqcn,mprdma,swift,ndp,bonato2025repsrecycledentropypacket,zeng2026connecting,jiang2026odin} to meet rapidly evolving workload requirements~\cite{dean2013tail,mapreduce,spark,memcache,li2020pytorch,zhao2023pytorch,gangidi2024rdma,lei2026fast,ma2026challenges}. Hardwiring these policies into the RDMA transport ASIC forces a multi-year cycle to tape out, test, and deploy new silicon~\cite{zhao2025white}. This cycle cannot keep pace with changing AI models, topologies, or congestion signals.

Emerging programmable RDMA engines shorten the cycle by accepting new algorithms after deployment, but their execution models and APIs tie policy code to a particular architecture, transport, or vendor~\cite{arashloo2020enabling,huang2025swcc,singhvi2025falcon,doca-pcc}. Software control planes offer a third option. CCP executes policy in a portable user-space control plane~\cite{narayan2018restructuring}, but it doesn't support offloading, and its microsecond-scale control cannot serve nanosecond-scale per-packet path decisions.

No common interface combines policy source portability with efficient execution across vendors and transports with different semantics. Operators need to develop a policy once and reuse it wherever the required feedback and controls exist. Vendors should implement each platform mapping once and optimize it independently of policy code. PCM provides this separation between policy development and platform integration (Figure~\ref{fig:poster}).

Programmable Congestion Management (PCM) is an open-source programming model for datacenter NIC policies. It separates policy development from platform integration through portable control objects and performance-optimized accessors. Developers express algorithms as event-driven \textit{handlers} over signals, triggers, controls, and private state. Platform backends generate \textit{shells} that map these objects to native state and CC/LB hooks, preserving handler source while allowing platform-specific optimization. This enables \textit{source compatibility}: developers reuse unchanged handler code across supported platforms by generating a shell for each target. Deployment parameters and behavioral validation remain target-specific.

Handler performance is as important as portability. At 1.5~GHz and 0.5 instructions per cycle, a core can execute only about 15 instructions per 4~KiB packet at 1.6~Tbit/s. Setting up a single function call can exhaust this budget. PCM closes this gap at three levels. First, shell generators compile NIC state accessors into native loads and stores to minimize abstraction overhead. Second, handlers for independent flows map transparently to parallel, energy-efficient NIC cores, increasing aggregate capacity. Third, hardware-aware signal accumulation preserves counts and feedback summaries while decoupling the execution of instruction-heavy or resource-limited handlers from event arrival rates. Together, these mechanisms let portable handlers approach bare-metal performance, react at nearly per-packet granularity when capacity permits, and consume accumulated feedback when policies tolerate coarser reactions.

To demonstrate PCM's expressivity, we implement eleven state-of-the-art datacenter CC and LB algorithms, most in fewer than 100 lines of C. We implement shell generators for four platforms. UET-htsim enables correctness checks and parameter tuning before deployment~\cite{ue-htsim}. Integrating PCM into UCCL tests portability to a CPU-based software RDMA transport~\cite{uccl_transport}. We support PCM offload on NVIDIA CX-8 and Microsoft 1P, two NIC architectures with different hardware RDMA transports.

Our evaluation answers four questions: What does the abstraction cost? What does policy reuse require? How does execution scale? What reaction rates can each platform sustain? Matched native baselines and targeted experiments reveal how arithmetic, feedback granularity, worker placement, and flow-state footprint shape the performance and behavior of portable PCM policies.

\hfill \break
\textit{We make the following contributions:}
\begin{enumerate}[leftmargin=1.2em, itemsep=0em]
    \item \textbf{Design requirements:} We derive expressivity, portability, and performance requirements from workload needs and existing mechanisms.
    \item \textbf{A unified, efficient programming model:} We define hardware-friendly abstractions for portable NIC policies with per-event and coarser reactions.
    \item \textbf{Multi-platform validation:} We demonstrate eleven policies, four platform integrations, and full CC offload.
    \item \textbf{Ultra Ethernet standardization impact:} Our work informed ongoing standardization of programmable congestion management.
\end{enumerate}
\newpage
\section{Motivation}\label{sec:motivation}

Three use cases motivate reusable policies across NICs.

\begingroup
\begin{figure}[h]
  \centering
  \captionsetup{skip=2pt}
  \captionsetup[subfigure]{skip=1pt}
  \begin{subfigure}[t]{0.49\linewidth}
    \centering
    \includegraphics[width=\linewidth]{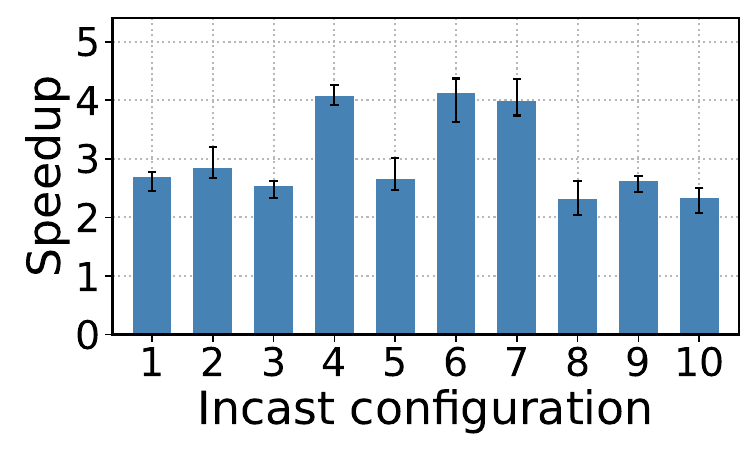}
    \caption{Incast-aware CWND tuning.}
    \label{fig:motivation_cwnd_tuning}
  \end{subfigure}\hfill
  \begin{subfigure}[t]{0.49\linewidth}
    \centering
    \includegraphics[width=\linewidth]{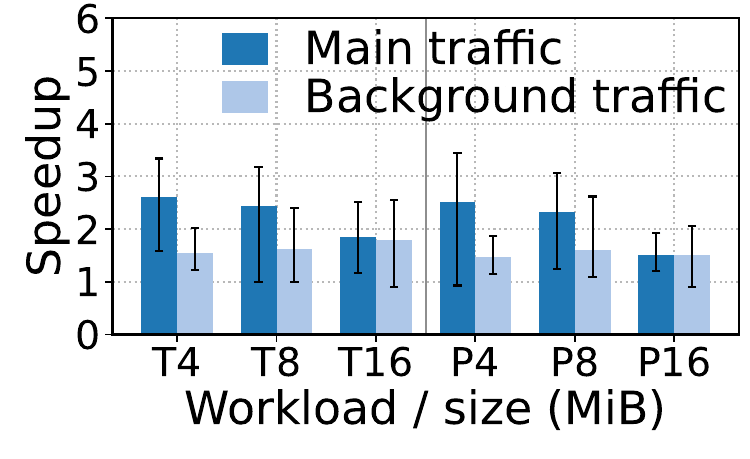}
    \caption{Mixed LB policies.}
    \label{fig:motivation_lb_mix}
  \end{subfigure}
  \caption{Maximum flow completion time (FCT) speedup. T/P: tornado/permutation. Bars show means over ten endpoint mappings; whiskers show min--max.}
  \label{fig:motivation_policies}
\end{figure}
\endgroup

\begin{table*}[!tb]
\centering
\small
\setlength{\tabcolsep}{3pt}
\resizebox{\textwidth}{!}{
\begin{tabular}{@{}llllll@{}}
\toprule
\textbf{Mechanism} & \textbf{Policy scope} & \textbf{Execution} & \textbf{Event handling} & \textbf{Integration requirements} & \textbf{Open source} \\
\midrule
Fixed NICs~\cite{de2020depth} & CC/LB & Sync & Per-event & ASIC & No \\
Tonic~\cite{arashloo2020enabling} & CC & Sync/Async & Per-event/periodic & FPGA, Verilog & Yes \\
AMD Pensando~\cite{amd-nic} & CC/LB & Not reported & Not reported & Pensando pipeline, P4-like language & No \\
Broadcom Thor~\cite{broadcom-nic} & CC/LB & Not reported & Not reported & Thor pipeline, C-like NPL language & Spec only \\
SwCC~\cite{huang2025swcc} & CC & Sync & Per-event & RoCEv2, SwCC RDMA engine, C & Yes \\
DOCA PCC~\cite{doca-pcc} & CC & Async & Event-specific HW coalescing & RoCEv2, NVIDIA DPA, C & Samples \\
Falcon~\cite{singhvi2025falcon} & CC/LB & Async & Event-specific HW coalescing & Falcon PDL/RUE & Spec only \\
Linux eBPF CC~\cite{hinz2023tcp} & CC & Sync & Per-event & Linux networking stack & Yes \\
UCCL~\cite{uccl_transport} & CC/LB & Sync & Per-event & Verbs, C++ & Yes \\
Presto~\cite{shashidhara2026presto} & CC & Async & Periodic & P4, C++ & Yes \\
CCP~\cite{narayan2018restructuring} & CC & Async & Arbitrary SW runtime filters & CC hooks, CCP DSL, Rust & Yes \\
PCM & CC/LB & Sync/Async & min/max/sum/last event accumulation & CC/LB hooks, C & Yes \\
\bottomrule
\end{tabular}
}
\caption{Sync/Async: policies run within/independently of the datapath. Integration requirements identify the datapath interfaces and execution support required by each mechanism.}
\label{tab:mechanism-comparison}
\end{table*}

\paragraph{Incast-aware CC tuning:}
PCM allows operators to explicitly tune policies for workloads and NICs. Even moderate incast can collapse throughput and raise tail latency in collectives, storage, analytics, and caching~\cite{dean2013tail,dcqcn,mapreduce,spark,memcache}. BDP-sized initial windows suit long, well-balanced flows such as all-to-all collectives, but cause packet loss at shallow-buffered ports under incast before CC converges. Figure~\ref{fig:motivation_cwnd_tuning} compares BDP-scale and incast-tuned smaller initial windows for the MP-RDMA policy in the packet-level Ultra Ethernet transport simulator (UET-htsim). Ten configurations (19--60 senders, 150--500~KB messages) use a 1024-node, two-tier 400~Gbit/s fat tree and ten shared random endpoint mappings each. We observe a $3\times$ speedup of maximum FCT. The tuned policy avoids 98.6\% of baseline packet drops.

\paragraph{Mixed LB policies:}
PCM benefits scenarios where applications impose conflicting requirements on packet delivery. Adaptive packet spraying lets AI collectives use all fabric paths' bandwidth, while co-deployed legacy MPI and some RDMA applications need in-order delivery~\cite{hoefler2025ultra,gangidi2024rdma,sdr}. PCM lets operators combine ECMP~\cite{ecmp} for legacy applications with adaptive schemes such as REPS~\cite{bonato2025repsrecycledentropypacket} for AI traffic. Figure~\ref{fig:motivation_lb_mix} compares per-flow coexistence of REPS and native ECMP against cluster-wide ECMP~\cite{ecmp,bonato2025repsrecycledentropypacket} in a simulated 128-node, two-tier 400~Gbit/s fat tree. Tornado and permutation workloads induce core congestion with 4, 8, and 16~MiB messages and ten random mappings per configuration~\cite{olteanu2022edge, bonato2025repsrecycledentropypacket}. REPS controls 115 of 128 flows. ECMP serves the 13 background flows, avoiding multipath-induced reordering. Compared to cluster-wide ECMP, REPS+ECMP improves maximum FCT by $1.51$--$2.61\times$ for main traffic and $1.48$--$1.80\times$ for background.

\paragraph{Domain-specific CCs:}
PCM enables endpoints to deploy custom CC policies per flow. This is particularly useful in heterogeneous multi-datacenter training~\cite{geminiTOn}. Operators typically deploy different policies depending on the traffic's destination: within a datacenter, low-latency schemes like Swift~\cite{swift}; across datacenters, delay-tolerant schemes like BBR~\cite{bbr,uno}.

\section{Related Work}\label{sec:mechanism-study}

Supporting workload-specific congestion management requires both programmable policies and reusable platform integration. Table~\ref{tab:mechanism-comparison} compares existing mechanisms for NIC congestion management. They bind policy code to hardware, interfaces, or host runtimes. PCM addresses this gap through CC/LB source portability across substrates, defined signal and control semantics, and native execution on embedded NIC cores. This combination extends policy reuse from simulation to NIC deployment.

\paragraph{Hardware and transport dependencies:}
HotCocoa provides programmable hardware abstractions for CC logic in Verilog~\cite{arashloo2017hotcocoa}. Tonic extends this approach to a 100~Gbit/s FPGA design, with Verilog templates for incoming-packet logic and periodic CC updates~\cite{arashloo2020enabling}. It processes relevant events every cycle, but policy changes require synthesis, timing closure, and target-specific verification. AMD Pensando and Broadcom Thor Ultra similarly tie P4-like pipeline code to vendor compilers, stages, metadata, and actions.

NVIDIA DOCA PCC compiles C callbacks for the ConnectX DPA, with ABI-specific flow contexts, events, algorithm attributes, and results~\cite{doca-pcc,zhao2025white}. It can coalesce events into aggregate counters. The DOCA PCC interface exposes RoCEv2 and RDMA engine control events and RTT probes, and provides nanosecond RTTs and fixed-point arithmetic helpers. 

Google's Falcon transport specifies an asynchronous engine that consumes ACK-, NACK-, and retransmission-triggered events and returns CC and path-selection controls~\cite{singhvi2025falcon}. The Falcon datapath filters control events according to queue occupancy, elapsed time, and acknowledged volume, while accumulating ACK counts between engine updates. These NICs expose native transport event formats, state layouts, and controls to policy code.

\paragraph{Programming language and runtime dependencies:}
Linux TCP loads telemetry-driven CC handlers as eBPF bytecode without rebuilding the kernel. Their C code depends on kernel callbacks (tcp\_congestion\_ops), helpers, and socket state, preventing offload to RDMA NICs~\cite{hinz2023tcp}. Presto's Tofino~2 RMT datapath implements TCP, aggregates acknowledged/ECN-marked bytes and duplicate ACKs, and enforces rates~\cite{shashidhara2026presto}. Its off-path C++ control plane runs DCTCP or Timely and updates rate tables, with configured minimum intervals of 50~$\mu$s between polls and 80~$\mu$s between datapath-generated credit messages. These policies retain dependencies on the P4 datapath and C++ control plane. UCCL implements GPU transport pacing, loss recovery, CC, and path selection over multi-packet chunks striped across RDMA queue pairs using Verbs UC Write~\cite{uccl_transport}. Portability across Verbs providers still leaves this logic tied to UCCL's classes and path APIs. All three systems require interface changes for deployment on other substrates. For example, Presto's and UCCL's C++ policies cannot compile with the DOCA PCC C11 compiler~\cite{doca-pcc}.

\begin{figure*}[t]
  \centering
  \includegraphics[width=\textwidth]{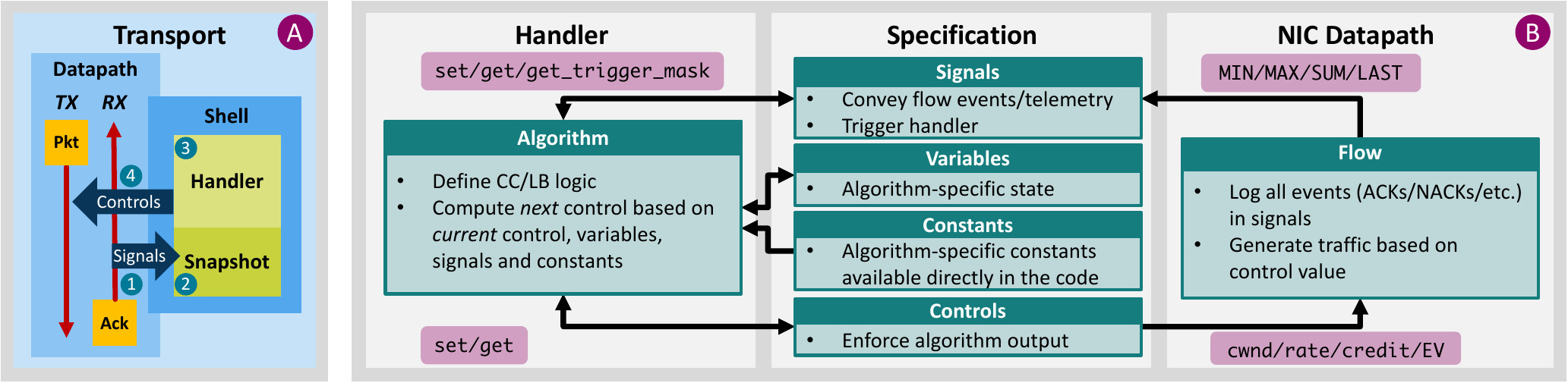}
  \caption{PCM separates a portable handler from the transport datapath. (A) A shell accumulates events, evaluates triggers, constructs a snapshot, executes the handler, and commits its outputs. (B) The specification names control objects, while each shell maps them to native mechanisms.}
  \label{fig:programming-model}
\end{figure*}

\paragraph{Host control plane delay:}
CCP preserves CC source across different host datapaths and protocols, including Linux TCP, QUIC, and mTCP~\cite{narayan2018restructuring}. It extends event coalescing with programmable filters and reporting rules. Policies can request per-event reports when the control-plane agent keeps pace or aggregate measurements before reporting. CCP interprets these rules as DSL programs on each ACK or timeout to accumulate measurements, select reports, and update controls. Portus, CCP's Rust runtime, runs the CC algorithm in a separate user-space process. It delivers reports to that algorithm and returns its control updates through POSIX IPC. On our Intel Xeon Gold 6140, CCP's NewReno cycle takes $10~\mu$s. A single Linux kernel IPC call takes about 3~$\mu$s on this system, the runtime-interpreted DSL takes 350~ns, and NewReno calculation takes 300~ns. At 1.6~Tbit/s, a 4~KiB packet arrives every 20~ns. CCP's filters and control plane round trip exceed this deadline by orders of magnitude. While some sender-side CC algorithms tolerate coarse control, others, such as MP-RDMA, target nearly per-packet reactions~\cite{mprdma}. Receiver-side CC likewise requires fine-grained credit allocation and pacing~\cite{olteanu2022edge}, while per-packet LB requires a path decision for each transmission~\cite{bonato2025repsrecycledentropypacket}. Host round trips for fresh per-packet decisions stall transmissions or leave the datapath using stale state during transient bursts and path congestion.

\section{Design Requirements}
\label{sec:pcm-requirements}

The use cases and NIC mechanisms discussed in Sections~\ref{sec:motivation} and~\ref{sec:mechanism-study} motivate three requirements for PCM.

\paragraph{Expressivity:}
A common programming model needs to express sender-side CC, receiver-side credit control, and stateful LB to accommodate diverse workload requirements.

\paragraph{Portability:}
Maintaining separate simulator and NIC implementations requires repeated integration work and creates opportunities for bugs. Policy reuse therefore requires separating handler logic from native APIs, state layouts, measurement units, and arithmetic implementations. We target \textit{source compatibility}: unchanged policy code across supported backends, with deployment-specific parameters and numerical precision. Feedback granularity, arithmetic, scheduling, and native and approximated controls can still affect behavior, so each deployment requires \textit{behavioral validation}. A backend must reject a handler if the platform cannot provide a required signal or enforce a required control.

\paragraph{Performance:}
Terabit rates demand low execution overhead and sufficient aggregate processing capacity. Portable handlers should approach native execution performance and scale across independent flows. Platforms also expose feedback at different granularities: DOCA PCC and Falcon can aggregate feedback before invoking policy code, while UCCL delivers ACKs for multi-packet chunks. PCM should exploit native execution and feedback mechanisms to minimize overhead and increase processing capacity. It should support per-event execution when the platform can sustain it and accumulated feedback when deployment-specific validation shows that the policy tolerates coarser reactions.

Section~\ref{sec:evaluation} examines these requirements through algorithm coverage, policy reuse across transports, and measurements of execution cost, goodput, and scaling.

\section{PCM Programming Model}~\label{sec:design}

We introduce a programming model to express congestion management algorithms as portable, event-driven \textit{handlers}.
A PCM handler is a stateful function that operates on per-flow state independently of hardware details.
Each handler context belongs to one flow.
Developers may compose modular logic by associating a flow with multiple handlers whose controls do not overlap. For example, separate handlers can implement load balancing and congestion control.
Matching flow events to a handler context is outside PCM's scope. It depends on the NIC transport, traffic-steering engine, and transport addressing.

\begin{figure*}[!tb]
  \centering
  \includegraphics[width=\textwidth]{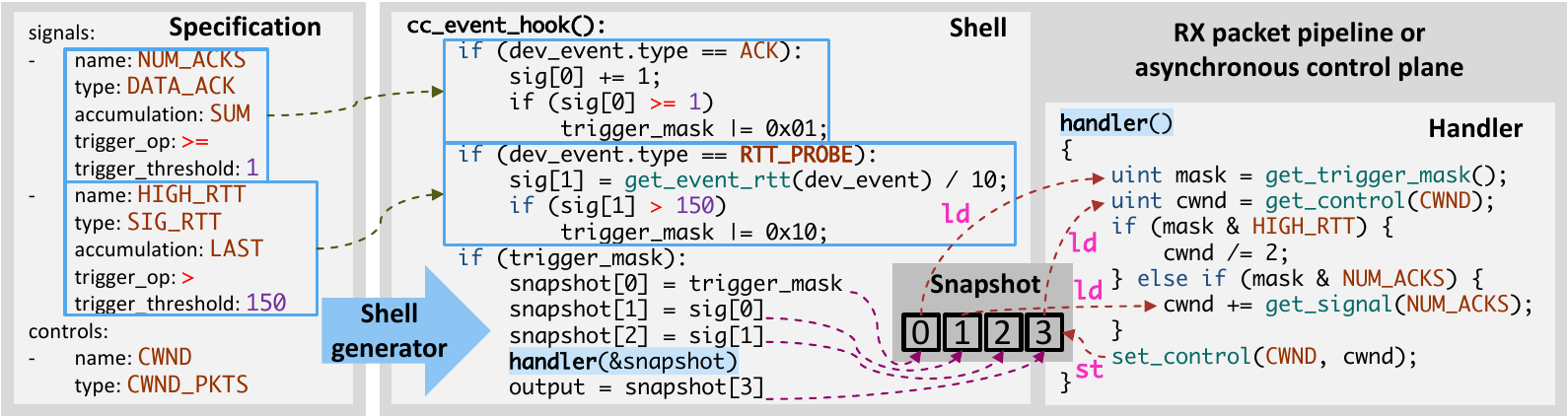}
  \caption{Shell generation from a specification and a handler. The generator emits constant object identifiers, the snapshot layout, and the required accumulation, trigger, snapshot, and commit operations. The compiler folds accessors into native loads and stores, eliminating runtime lookup and mask conversion.}
  \label{fig:shell-generation}
\end{figure*}

\subsection{Unified Datapath Abstractions}

Figure~\ref{fig:programming-model} shows how PCM separates congestion management policy from platform-specific event management and control enforcement.
A handler uses \texttt{get} and \texttt{set} accessors to read signals, update controls, and maintain private state.
These accessors provide a safe API that hides each object's internal representation.

\paragraph{Controls:}
A \textit{control} specifies a transport knob, such as a window, rate, path entropy, or cumulative credit.
After execution, the handler returns its controls to the transport.

\paragraph{Signals:}
A \textit{signal} exposes datapath telemetry or live state.
Examples include counts of ACKs, NACKs, RTOs, and ECN marks, as well as RTT, elapsed time, queue backlog, and path identity. Each signal selects an accumulator: \texttt{SUM} for counts, \texttt{MIN}/\texttt{MAX} for extrema, or \texttt{LAST} for the newest value.
Accumulator storage is independent of event count, and several named accumulators can share one telemetry source.

\paragraph{Triggers:}
PCM invokes handlers through a \textit{triggering mechanism} driven by accumulated flow events.

A trigger uses a particular signal value to determine when a handler should run.
The backend combines all fired triggers into one mask.
The handler reads the mask to select its branches.
Each handler requires at least one triggering signal.

Each trigger evaluates a single-instruction condition against the accumulated value and a static \textit{threshold}.
The comparison supports several modes.
\texttt{RAW} compares the accumulated value with the threshold.
\texttt{DELTA} compares the change since that trigger's last satisfied check.
\texttt{MAGNITUDE} compares the absolute change from the same baseline. For example, an ACK counter with \texttt{DELTA >= 1} permits per-packet reactions when the shell keeps pace.
If the control law permits, a higher threshold allows less frequent reactions.

\paragraph{Variables and constants:}
\textit{Variables} hold persistent private state, including signed and unsigned scalars and fixed-size arrays.
Developers can choose floating-point or fixed-point representations for fractional variables at compile time.
\textit{Constants} define state that remains immutable throughout a flow, such as base RTT and MTU.
The compiler resolves them at compile time to optimize calculations and eliminate unnecessary loads.

\subsection{Handler Compilation}
\label{sec:datapath-specification}
\label{sec:code-generation}

The specification lists all objects used by the handler and their properties.
The C source and specification form a handler package.
Figure~\ref{fig:shell-generation} shows how a compiler combines this package with a platform backend API to produce a \textit{shell}.

\paragraph{Object generation:} The shell translates datapath events to signals and maps controls to native mechanisms.
For each specified object, the compiler emits logic that implements its semantics.

Generated code for \texttt{HIGH\_RTT} in the example in Figure~\ref{fig:shell-generation} branches on the native RTT-probe event and checks whether the converted RTT exceeds 150 PCM ticks ($15~\mu$s).

\paragraph{Object representation:} Shells can normalize or approximate exposed feedback and controls.
The example in Figure~\ref{fig:shell-generation} converts the probe from 10~ns platform ticks to 100~ns PCM ticks.
Some handlers evaluated in Section~\ref{sec:evaluation} use byte counts.
On ConnectX, the backend API delivers only cumulative ACK counts. The shell estimates acknowledged bytes by multiplying that count by the configured MTU.
However, shells cannot recover unavailable information or capabilities.
The compiler checks interface support before deployment.
For example, our ConnectX backend rejects LB handlers because its backend interface lacks path selection.

\paragraph{Source compatibility:}
Successful shell generation from the handler package establishes \textit{source compatibility}.
Developers validate handler behavior in simulation and deployment with the available signals, constants, and numerical precision.
Vendors validate event accounting, units, state placement, scheduling, and native control mappings.

\paragraph{Compile-time optimizations:}
Separating the handler from its specification enables compile-time optimizations (Figure~\ref{fig:shell-generation}).
We exploit these opportunities in all our shell compilers (Section~\ref{sec:implementation}).
The compiler can use the specification to set up a static memory layout for handler objects, fold object accesses into native loads and stores, and map accumulator updates to hardware primitives, such as ConnectX's event coalescing or atomic instructions.
On both AArch64 and RISC-V, each signal accumulation takes one (SUM, LAST) or two (MIN, MAX) instructions with operands in registers.

\subsection{Handler Execution Modes}
\label{sec:exec-modes}
PCM defines two execution modes relative to flow events.

In synchronous mode, the shell completes the execution cycle in Figure~\ref{fig:programming-model}A on the flow's event path.
Signal updates and execution do not overlap, so the shell can use non-atomic context storage.

In asynchronous mode, a separate control plane evaluates triggers and runs handlers independently of datapath progress.
The shell exchanges signals and controls between the data and control planes.
Our asynchronous CPU shell connects the datapath API to accumulators and controls through CPU atomics.
On ConnectX, the closed-source PCC runtime exchanges events and control results between the RDMA datapath and the DOCA PCC flow context. This context runs on a statically assigned hardware thread~\cite{doca-pcc}.

\paragraph{Snapshots:}
\textit{Snapshots} preserve \texttt{get}/\texttt{set} signal semantics across both execution modes (Figure~\ref{fig:shell-generation}).
Datapath updates cannot change the captured values during execution. Handler writes remain local until commit.

Some transports may update multiple signals per event, such as an ACK carrying a cumulative acknowledged count and an RTT sample.
Asynchronous shells may need to update individual signal accumulators one by one, as our CPU shell does through atomics.
Locking all accumulators while copying a snapshot could stall both planes.
To avoid these stalls, PCM permits relaxed snapshot delivery.
Each signal preserves accumulation order, but the handler may read feedback from different update epochs.
For general bounded skew, we propose letting policies specify a maximum and shells tag updates with sequence numbers to discard snapshots that exceed it.
Our evaluated asynchronous CPU shell implements relaxed and zero-skew snapshots.

\subsection{Real-World Handler Examples}
\label{sec:api-examples}
We show key PCM interface concepts in excerpts from three real handlers in Table~\ref{tab:algorithm-coverage}.

\paragraph{Sender-side CC:}
Listing~\ref{lst:cc} illustrates ACK feedback consumption by an MP-RDMA-style sender CC~\cite{mprdma}. We provide the corresponding handler's specification in Appendix~\ref{sec:spec-appendix}, Listing~\ref{lst:cc-yaml}. It declares a cumulative ACK-byte counter, a \texttt{DELTA >= 1} trigger, and a private baseline~\texttt{PREV\_ACKED\_BYTES} variable.
On new ACKs, the handler subtracts the baseline to obtain newly acknowledged bytes. It then advances the baseline and writes its window decision through a control.

\begin{lstlisting}[style=pcmcode,float=tb,caption={ACK/NACK-based sender CC.},label={lst:cc}]
uint cwnd = get_control(CWND);
if (get_trigger_mask() & NACKED_BYTES) {
  // decrease logic
} else if (get_trigger_mask() & ACKED_BYTES) {
  uint cur_aked = get_signal(ACKED_BYTES);
  uint diff = cur_aked - get_var(PREV_ACKED_BYTES);
  cwnd += diff * MTU / cwnd;  // Additive increase.
  cwnd = max(MTU, cwnd);
  set_var(PREV_ACKED_BYTES, cur_aked);
}
set_control(CWND, cwnd);
\end{lstlisting}

\begin{lstlisting}[style=pcmcode,float=tb,caption={EQDS grant step for unsatisfied sender demand.},label={lst:receiver-cc}]
if (get_trigger_mask() & (PULL_TARGET | NEEDS_MORE)) {
  uint target = get_signal(PULL_TARGET);
  uint pull = get_control(PULL_NO);
  if (target > pull)
    pull += min(CREDITS_PER_PULL, target - pull);
  set_control(PULL_NO, pull);
  set_signal(NEEDS_MORE, pull < target ? 1 : 0);
}
\end{lstlisting}

\begin{lstlisting}[style=pcmcode,float=tb,caption={REPS ACK handling: cache "good" EVs.},label={lst:lb}]
uint mask = get_trigger_mask();
uint cur_ecn = get_signal(NUM_ECN);
uint is_ecn = cur_ecn - get_var(PREV_ECN);
if (mask & NUM_ACK) {
  set_var(PREV_ECN, cur_ecn);
  if (!is_ecn) {
    uint head = get_var(EVC_HEAD_IDX);
    uint count = get_var(EVC_NUM_VALID_EVS);
    set_arr(EVC, head, get_signal(ACK_EV));
    if (!get_bitmap_idx(EVC_BITMAP, head)) {
      set_var(EVC_NUM_VALID_EVS, count + 1);
      set_bitmap_idx(EVC_BITMAP, head, 1);
    }
    head = (head + 1) & (EVC_SIZE - 1);
    set_var(EVC_HEAD_IDX, head);
  }
}
\end{lstlisting}

The shell's \texttt{DELTA} baseline determines when to invoke the handler. The private ACK baseline records the consumed feedback.
Branch order also matters: NACK represents negative feedback and takes precedence when both triggers fire.

\paragraph{Receiver-side CC:}
Our EQDS receiver~\cite{olteanu2022edge} illustrates how a handler paces an individual sender.
It reads a cumulative credit target and publishes a cumulative pull number to grant credit.
\texttt{PULL\_TARGET} uses a \texttt{MAX} accumulator and a \texttt{MAGNITUDE >= 1} trigger, retaining the largest requested target. A target increase starts the handler.

Listing~\ref{lst:receiver-cc} grants up to \texttt{CREDITS\_PER\_PULL} to a flow. It caps the new pull number at the target and publishes it through \texttt{PULL\_NO}. If demand remains, the handler sets \texttt{NEEDS\_MORE}, a loopback signal internal to the handler and shell.
This binary flag requests further service through a \texttt{RAW >= 1} trigger.
The transport schedules handlers, arbitrates flows, and paces pulls. After serving other flows, it can invoke this handler again on the loopback trigger.

\paragraph{LB:}
REPS uses static arrays to cache healthy, uncongested paths for later transmissions~\cite{bonato2025repsrecycledentropypacket}.
Listing~\ref{lst:lb} shows how it populates an EV cache on the ACK RX path. The handler derives the count of new ECN marks from a private baseline and caches the last reported path only when this count is zero.
Indexed access through \texttt{set\_arr} hides the array's storage address. A power-of-two cache size bounds the masked index.
\texttt{ACK\_EV} uses \texttt{LAST} to identify the path associated with the feedback.
The handler's TX path (not shown) finds an EV in the cache and returns it to the transport. The transport could then add that EV to its table of active EVs.

This example also exposes a semantic limit of accumulation.
A \texttt{LAST} path signal exposes one identity and cannot preserve all ACK-to-path associations in a coalesced batch.
Exact per-packet cache updates therefore require coherent event delivery and sufficient compute resources, as in the synchronous simulator validation.

\begin{table*}[t]
\centering
\small
\setlength{\tabcolsep}{3pt}
\setlength{\arrayrulewidth}{0.2pt}
\resizebox{\textwidth}{!}{
\begin{tabular}{@{}lllll@{}}
\toprule
\textbf{System} & \textbf{Execution} & \textbf{Shell} & \textbf{Transport} & \textbf{Testbed} \\
\midrule
UET-htsim & Sync / Async & CPU Reference Shell & Ultra Ethernet &
\begin{tabular}[t]{@{}l@{}}1,024-host, three-tier fat tree, 100~Gbit/s links, 4,150~B packets \end{tabular} \\
CPU microbenchmark & Async & CPU Reference Shell & Synthetic trace & Dedicated Intel Xeon Gold 6140 core as generator and consumer \\
\cline{5-5}
UCCL & Sync / Async & CPU Reference Shell & UCCL UC Verbs &
\multirow{2}{*}{\begin{tabular}[c]{@{}l@{}}Two AMD EPYC 9124 servers, each with two 800~Gbit/s ConnectX-8 NICs,\\
sixteen 100~Gbit/s planes per server, Spectrum-5 multi-plane, two-tier fabric\end{tabular}} \\
CX-8 & Async & DOCA PCC Shell & RoCEv2 & \\
\cline{5-5}
DPA microbenchmark & Async & DOCA PCC Shell & Synthetic trace & AMD EPYC 9124 core as generator and ConnectX-8 DPA threads as consumers \\
M-1P & Async & 1P NIC Shell & Custom RDMA & Two 2$\times$200~Gbit/s Microsoft 1P DPUs and an Arista DCS-7060X6 switch\\
\bottomrule
\end{tabular}
}
\caption{PCM platform shells and testbed systems.}
\label{tab:platform-mappings}
\end{table*}

\section{Implementation}~\label{sec:implementation}

Table~\ref{tab:platform-mappings} summarizes PCM mappings and testbeds.
Our platform integrations take approximately 600 (htsim, C++), 1000 (UCCL, C++), 2500 (Microsoft 1P, C/Python), and 2000 (CX-8 DOCA PCC, C/Python) lines of code.

\subsection{CPU Reference Shell}

Our reference CPU shell is implemented in C++ and supports integration with UET-htsim and UCCL.

The handler specification is represented as a type tuple that instantiates a variadic template shell class. The instantiated shell class specializes a shell for a given handler. Fold expressions and \texttt{if constexpr} omit redundant checks.
The shell class can be specialized with signal storage for synchronous or asynchronous execution.
Synchronous storage uses ordinary C++ datatypes and operations.
Asynchronous storage uses an atomic fetch-add for \texttt{SUM}, a store for \texttt{LAST}, and compare-and-exchange for \texttt{MIN}/\texttt{MAX}.

We compile the shell into a separate shared object.
Datapaths load algorithms and use the shell factory to create per-flow shell instances. A virtual API flushes a datapath-input structure into signal accumulators. A mask selects feedback, byte counts, RTT, queue state, and path information to selectively apply feedback to the signal accumulators. Separate structure fields return window, rate, and entropy controls.
We compile handlers with Clang and LLVM~\cite{lattner2004llvm} and also provide optional custom passes to check handler safety, including detecting potentially unbounded execution paths.

\paragraph{UET-htsim:}
Simulator hooks map the simulator clock, ACK, NACK, ECN, LB and credit pull request events to the shell input structure.
Synchronous mode applies controls before window enforcement, path selection or credit generation.
We also extend the simulator's NIC scheduler to model an asynchronous control engine that executes handlers in a separate event chain.

\paragraph{UCCL:}
UCCL distributes flows across CPU threads that fragment buffers into multi-MTU chunks and provide reliable delivery over unreliable Verbs RDMA Writes.
UCCL defaults to a software CC module that runs synchronously on the chunk-processing path.
We connect it to our synchronous CPU shell.
Synchronous PCM maps ACK, NACK, and RTT feedback and applies the resulting window inside the transport callback. 
We also extend UCCL to support an asynchronous CC plane that serves all datapath threads from an independent thread.
The two planes communicate through the CPU reference shell's atomic storage.
The PCM worker scans contexts and checks triggers. It publishes window updates on separate atomic cache lines for independent datapath reads.

\subsection{ConnectX DOCA PCC Shell}
\label{sec:cx8-impl}

The ConnectX-8 shell offloads PCM to Datapath Accelerator (DPA) cores through the DOCA PCC API.
The DPA has 16 cores, each with 16 hardware threads.
The shell uses DOCA PCC callbacks that run asynchronously relative to the hardware RoCEv2 engine~\cite{doca-pcc}.
Each per-flow callback runs the shell and returns a rate and RTT-probe requests.
The shell generator uses the handler's specification to emit DOCA PCC glue code that is compiled using the DPA's DPACC compiler. This code wraps native SUM-based coalescing for TX-burst, ACK, NACK, and CNP events and handles RTT probes.

\paragraph{Per-flow state:}
DOCA PCC provides 48 application-context bytes per flow (RC QP).
This space holds a handler context pool index, output rate, outstanding bytes, probe state, RTT, and debugging metadata.
The first event atomically reserves a pool slot and initializes the flow's generated handler state.

\paragraph{DPA arithmetic:}
The DPA lacks floating-point hardware and supports only a few mathematical intrinsics in the fixed-point Q16.16 format.
To support Swift and CUBIC, we implement arithmetic, comparison, root, reciprocal, and inverse-square-root macros.

These macros emulate fixed-point arithmetic in software using the open-source fixedptc library.
The library supports 32- or 64-bit formats chosen at compile time. Our production implementation will replace these macros with a custom shell compiler pass that uses C arithmetic operators and a subset of the \texttt{math.h} API.

\paragraph{Native controls:}
The only datapath control knob that DOCA PCC supports is QP rate.
The shell exposes rate control to PCM handlers in bits per 100~ns tick. It maps this value to ConnectX's native Q20 integer encoding of the link-rate fraction.
For functional testing of purely window-based algorithms, the shell adapts Swift's native pacing method to approximate rate as \(CWND/\mathit{EWMA(RTT)}\), using an exponentially weighted moving average of RTT~\cite{swift,ue-htsim,uccl_transport}.
Native window support in future ConnectX generations would let the shell select the window output.

\subsection{Microsoft 1P NIC Shell}
We implement a shell generator for Microsoft 1P NICs that run a custom hardware RDMA transport.
The C shell runs as an asynchronous CC callback on 8 of the 192 hardware threads co-located with the RDMA engine. Like the DOCA PCC shell, it dispatches events to per-flow state and controls to native 1P APIs.

\section{Evaluation}
\label{sec:evaluation}

\textbf{We answer four research questions:}
\begin{enumerate}
\item \emph{What does PCM's abstraction cost?} 
\item \emph{How does PCM support policy validation and reuse across transports?}
\item \emph{How does PCM scale across datapaths?}
\item \emph{What rates do PCM CC and LB handlers sustain?} 
\end{enumerate}

\begin{figure}[]
  \centering
  \includegraphics[width=\linewidth]{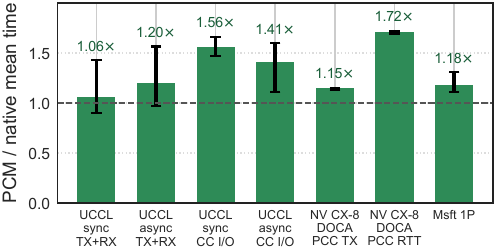}
  \caption{PCM mapping cost across three transports, averaged across 1--4 UCCL workers, 100--2000 ConnectX QPs, and 16 Microsoft 1P QPs. Whiskers show the minimum and maximum. UCCL CC I/O isolates CC cost, while TX+RX covers per-chunk work. In asynchronous UCCL runs, we measure the overhead incurred by UCCL datapath threads. The PCM control-plane worker runs on the dedicated SMT sibling of datapath thread 0, and its cost is excluded.}
  \label{fig:eval-overhead}
\end{figure}

\begin{figure*}[!t]
  \centering
  \begin{subfigure}[t]{0.24\textwidth}
    \centering
    \includegraphics[width=\linewidth]{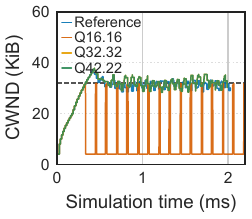}
    \caption{CWND numerical stability.}
    \label{fig:eval-swift-trace}
  \end{subfigure}\hfill
  \begin{subfigure}[t]{0.24\textwidth}
    \centering
    \includegraphics[width=\linewidth]{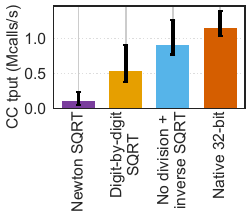}
    \caption{Arithmetic optimizations.}
    \label{fig:eval-swift-arithmetic}
  \end{subfigure}\hfill
  \begin{subfigure}[t]{0.24\textwidth}
    \centering
    \includegraphics[width=\linewidth]{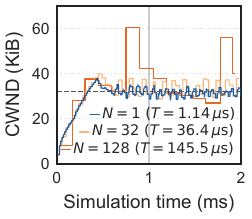}
    \caption{CWND fidelity under load.}
    \label{fig:eval-swift-async-cwnd}
  \end{subfigure}\hfill
  \begin{subfigure}[t]{0.24\textwidth}
    \centering
    \includegraphics[width=\linewidth]{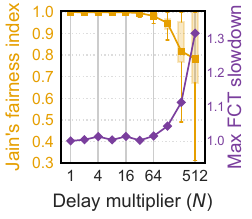}
    \caption{Fairness and slowdown.}
    \label{fig:eval-swift-schedule-fairness}
  \end{subfigure}
  \caption{PCM-Swift under a 3~MB 8:1 UET-htsim incast with senders 25~$\mu$s apart. Panel (a) shows Leaf~1 CWND for the synchronous floating-point reference and asynchronous fixed-point handlers delayed by $T_h\approx1.1~\mu$s. $T_h$ is the mean processing time of the optimized Q42.22 variant in panel (b). Panel (b) shows four DPA arithmetic variants: three in Q42.22 and one using CX-8's DPA-native Q16.16 format. Panels (c) and (d) use scheduling delays of $N T_h$.}
  \label{fig:eval-swift-debug}
    \vspace{-1em}
\end{figure*}

\subsection{Experimental Setup}
\label{sec:experimental-setup}

\paragraph{Live experiments scope:} We implement the eleven handlers in Table~\ref{tab:algorithm-coverage} across the platforms in Table~\ref{tab:platform-mappings} and validate all of them in UET-htsim. We test all seven sender-CC handlers under live traffic on UCCL, ConnectX, and Microsoft 1P. The UCCL and ConnectX tests use ACK/RTT-only feedback on a fabric without ECN or trimming.

\paragraph{CPU and ConnectX DPA microbenchmarks:} We implement a trace-replay microbenchmark to isolate performance at scales beyond our two-node, four-NIC testbed. A dedicated host CPU thread drives events. For CPU tests, it sends events through the CPU's atomic interconnect to an asynchronous reference shell on another hardware thread. For DPA tests, it sends events to DPA threads that poll host-DPA CQs and demultiplex completions into the DOCA PCC PCM shell. For a given handler, both DPA and CPU execute the PCM shell code that is executed for it when it serves real traffic.

When reporting microbenchmark results for sender-CCs, we use four event mixes and run each mix with three independent seeds. The ACK-only trace contains 100\% ECN-free ACKs. The ECN-heavy trace contains 90\% ECN-marked ACKs and 10\% unmarked ACKs. The loss-heavy trace contains 80\% loss events and 20\% unmarked ACKs. The mixed trace contains 20\% loss events, 40\% ECN-marked ACKs, and 40\% unmarked ACKs. On the DPA, we expand each synthetic trace event into native DOCA PCC events. Each loss event becomes two events: a TX burst and NACK. Each ACK becomes TX, RTT, and ACK events. For an ECN-marked ACK, we also add a CNP packet event to this sequence.

We use synthetic traces to measure peak execution capacity of each shell and its handler. Our evaluation of LB and receiver-CC handlers is limited to synthetic traces because neither CX-8 nor M-1P NICs expose public APIs for LB or receiver-CC hooks. For LB, both CPU and DPA microbenchmarks reuse the sender-CC traces on RX path hooks, and supply synthetic TX events that indicate one packet ready for transmission. For EQDS, the CPU microbenchmark uses four receiver-side pull-target traces in which 0\%, 25\%, 50\%, or 75\% of events advance the target by 128 credit units (two grant quanta), while the remaining increments are sampled uniformly from 1--127 credit units.

On our Intel Xeon testbed, the CPU generator thread saturates the sibling hardware thread on the same physical core at 35 million events/s (1.15~Tbit/s assuming 4~KiB packets). In our DPA microbenchmark, an AMD EPYC CPU saturates the DPA at 75 million events/s (2.5~Tbit/s).

\begin{table}[]
\centering
\resizebox{\columnwidth}{!}{
\begin{tabular}{llrr}
\textbf{Algorithm} & \textbf{Key feature} & \textbf{Size (B)} & \textbf{LoCs}\\
\toprule
NewReno~\cite{ietf-newreno} & Loss-based & 176 & 50 \\
CUBIC~\cite{ha2008cubic} & Loss-based & 240 & 95\\
DCTCP~\cite{dctcp} & Loss/ECN-based & 216 & 69 \\
MP-RDMA~\cite{mprdma} & ECN-based & 128 & 27 \\
Swift~\cite{swift} & RTT-based & 208 & 93\\
SMaRTT~\cite{smartt} & ECN/RTT/Trim hybrid & 264 & 104 \\
NSCC~\cite{uec-spec} & UET SMaRTT variant & 320 & 172\\
EQDS~\cite{olteanu2022edge} & Receiver-based & 232 & 91\\
\midrule
OPS~\cite{rps} & Oblivious per-packet & 96 & 38\\
REPS~\cite{bonato2025repsrecycledentropypacket} & Caches "good" paths & 232 & 70\\
STrack~\cite{le2024strack} & Tracks per-path congestion & 1240 & 75\\
\end{tabular}
}
\caption{CC and LB algorithms implemented in PCM.}
\label{tab:algorithm-coverage}
\end{table}

\subsection{Cost of Abstraction}
\label{sec:abstraction-cost}

Figure~\ref{fig:eval-overhead} compares built-in CC algorithms with their PCM reimplementations under live traffic on three platforms. The algorithms are Swift in UCCL, rtt\_template in DOCA PCC (CX-8), and a custom Microsoft 1P (M-1P) CC algorithm. The CX-8 and M-1P ports are excluded from Table~\ref{tab:algorithm-coverage}. For matched algorithms, we compare control behavior under the tested traffic conditions through CWND-adjustment distributions or rate trajectories. The asynchronous UCCL bars in Figure~\ref{fig:eval-overhead} measure only the added datapath work and exclude the control plane worker. A dedicated worker is a resource cost of asynchronous control plane design, not an intrinsic PCM requirement.

\paragraph{CPU-based transport:} In UCCL, one to four datapath threads perform 64~KiB chunk fragmentation. PCM increases the isolated cost of CC input and output by $\approx1.5\times$ in both modes. Native CC occupies 7.4--10.6\% of the TX+RX path. This path includes chunk transmissions and ACK handling. Overall TX+RX costs average up to $1.06\times$ and $1.20\times$ native execution for synchronous and asynchronous PCM, respectively. Notably, native UCCL and both PCM extensions reach the same 1.56~Tbit/s practical-goodput rate with four datapath threads.

\paragraph{NIC offloading:} Our CX-8 testbed uses the default DOCA PCC allocation of 48 of the DPA's 256 hardware threads, approximately one-fifth of the available threads.
For TX-burst callbacks that issue RTT probes, PCM increases the cost from 214 to 246~ns ($1.15\times$). 
The cost of RTT-probe-arrival callbacks that execute the handler increases from 392 to 676~ns ($1.72\times$).
We observe roughly 0.7 RTT-probe arrivals per TX-burst callback.
This ratio corresponds to about one RTT callback per 1.4 TX callbacks.
The combined overhead is $1.48\times$.
On the M-1P testbed, PCM increases the ACK-reaction cost by $1.18\times$. The shell runs on 8 of the NIC's 192 hardware threads, one twenty-fourth of the available threads.

\begin{figure*}[!t]
  \centering
  \begin{subfigure}[t]{0.24\textwidth}
    \centering
    \includegraphics[width=\linewidth]{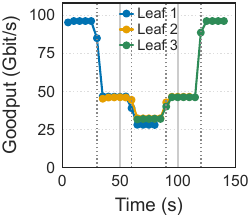}
    \caption{64~KiB RDMA Write.}
    \label{fig:eval-connectx-goodput}
  \end{subfigure}\hfill
  \begin{subfigure}[t]{0.24\textwidth}
    \centering
    \includegraphics[width=\linewidth]{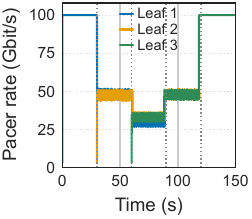}
    \caption{Pacer rate evolution.}
    \label{fig:eval-connectx-rate}
  \end{subfigure}\hfill
  \begin{subfigure}[t]{0.24\textwidth}
    \centering
    \includegraphics[width=\linewidth]{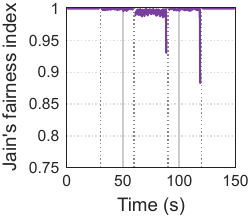}
    \caption{Pacer rate fairness.}
    \label{fig:eval-connectx-fairness}
  \end{subfigure}\hfill
  \begin{subfigure}[t]{0.24\textwidth}
    \centering
    \includegraphics[width=\linewidth]{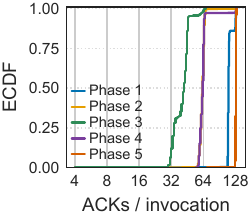}
    \caption{Coalescing during logging.}
    \label{fig:eval-connectx-coalescing}
  \end{subfigure}
  \caption{PCM-Swift on ConnectX-8 DPA controls end-to-end 3:1 incast on one 100~Gbit/s plane. Three leaves issue back-to-back 64~KiB RDMA Writes in five phases, starting 30 seconds apart and stopping in order.}
  \label{fig:eval-connectx-swift}
  \vspace{-1em}
\end{figure*}

\subsection{PCM-Swift Case Study}
\label{sec:eval-swift}

We follow the PCM workflow in Figure~\ref{fig:poster}: validate Swift's numerical representation in UET-htsim, optimize its computation, and test its scheduling tolerance. We then deploy the same C policy source on three live platforms, changing deployment constants such as MTU and BDP in the specification. Platform shells provide target-specific arithmetic and native control mappings. We validate the resulting deployments through goodput, control behavior, and fairness.

We use Swift for end-to-end behavioral validation for two reasons. First, it performs one of the most complex computations among the supported algorithms. Its target-RTT and estimated-RTT computations involve division and square-root operations and together account for $\approx90\%$ of the per-ACK processing cost. This makes Swift a demanding test of arithmetic performance on our NIC offloading platforms. Second, Swift requires only ACK and RTT feedback, matching the signals available in our Spectrum-5 fabric, and natively supports pacing rate control, the only datapath control available through ConnectX's DOCA PCC.

\paragraph{Step 1: Correctness testing.}

The CX-8 DPA is one of our offload targets. It lacks floating-point hardware, so we test fixed-point alternatives. The DPA-native 32-bit Q16.16 fixed-point data type has 16 integer and 16 fractional bits. In Figure~\ref{fig:eval-swift-trace}, we establish a stable fixed-point representation that preserves the tested numerical behavior. Swift's CWND output with Q16.16 is distorted.

The Q16.16 variant's CWND oscillates between its one-packet initial window and approximately 32~KiB. This variant takes 8.76~ms to complete the flows, compared with 2.17~ms for the reference. It also causes 1030 retransmissions. The 64-bit Q32.32 and Q42.22 variants complete all flows within 0.3\% of the reference's maximum FCT. They also achieve mean CWND fairness within 0.001 of the reference.

\paragraph{Step 2: Performance tuning.}
Our initial reference PCM-Swift computation follows the built-in htsim implementation. The naive Swift computation uses a floating-point square-root (sqrt) calculation to estimate the target delay.

\emph{Shell optimization.} The fixedptc library's baseline Newton approximation, which implements sqrt in the ConnectX DPA shell, sustains only 0.105 million calls/s (\mbox{Mcalls/s}) per hardware thread (Figure~\ref{fig:eval-swift-arithmetic}). An optimized digit-by-digit sqrt algorithm yields 0.532 \mbox{Mcalls/s}.

\emph{Handler optimization.} We further improve performance through algebraic transformations of Swift arithmetic. We replace divisions and the square root in Swift's target-delay computation with multiply-only reciprocal and inverse-square-root operations. The transformed handler exceeds 0.9 \mbox{Mcalls/s} per DPA thread, with a mean processing time of approximately $1.1~\mu$s per input event.

This example demonstrates Swift policy reuse during optimization: after each refinement, we revalidated the policy in the simulator as described in Step 1 and then benchmarked it unchanged on the DPA.

\paragraph{Step 3: Scheduling tolerance bound.}

We characterize Swift's sensitivity to asynchronous scheduling delay under the selected incast workload. After each invocation, the simulator waits $T=N T_h$ before invoking the flow's handler again, where $T_h\approx1.1~\mu$s is the mean processing time for optimized PCM-Swift on the CX-8 DPA. The multiplier $N$ models reduced handler availability on a shared DPA thread. We impose this delay directly without adding competing traffic flows. During the delay, the transport continues processing packets, and the shell accumulates the flow's ACK and RTT feedback for the next invocation.

\emph{Workload-specific tolerance.} Figure~\ref{fig:eval-swift-async-cwnd} shows Flow~1 under delay multipliers $N=1$, $32$, and $128$. Larger scheduling delays produce coarser CWND updates. Figure~\ref{fig:eval-swift-schedule-fairness} quantifies the resulting changes in fairness and completion time. Median CWND fairness stays at or above 0.99 from $N=1$ through $N=32$. It degrades to 0.818 at $N=256$ and 0.782 at $N=512$, corresponding to reaction delays of 291 and 582~$\mu$s. Maximum FCT increases by $1.11\times$ and $1.32\times$ relative to the $N=1$ result, respectively. At $N=32$, the 36.4~$\mu$s reaction interval spans the transmission time of 111 packets at 100~Gbit/s with a 4~KiB MTU.

Falcon's FAE study also reports sensitivity to reaction delays: flow completion time increases when the control reaction delay exceeds $32~\mu$s at 200~Gbit/s~\cite{singhvi2025falcon}. Our simulated sensitivity also provides context for Presto's $50~\mu$s control-plane polling period~\cite{shashidhara2026presto}.

\begin{figure*}[!t]
  \centering
  \begin{subfigure}[t]{0.24\textwidth}
    \centering
    \includegraphics[width=\linewidth]{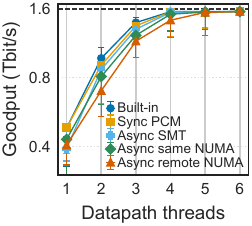}
    \caption{1~MB RDMA Write.}
    \label{fig:eval-uccl-throughput}
  \end{subfigure}\hfill
  \begin{subfigure}[t]{0.24\textwidth}
    \centering
    \includegraphics[width=\linewidth]{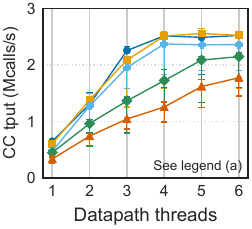}
    \caption{PCM handler-call rate.}
    \label{fig:eval-uccl-placement}
  \end{subfigure}\hfill
  \begin{subfigure}[t]{0.24\textwidth}
    \centering
    \includegraphics[width=\linewidth]{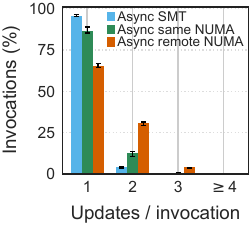}
    \caption{ACK SUM signal coalescing.}
    \label{fig:eval-uccl-coalescing}
  \end{subfigure}\hfill
  \begin{subfigure}[t]{0.24\textwidth}
    \centering
    \includegraphics[width=\linewidth]{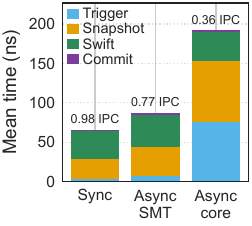}
    \caption{Worker shell breakdown.}
    \label{fig:eval-uccl-profile}
  \end{subfigure}
    \caption{PCM-Swift in CPU-based UCCL across sixteen ConnectX-8 100~Gbit/s planes (1.6~Tbit/s) and 16 flows on an AMD EPYC 9124 CPU. Panel (a) shows the median and extrema from three runs. Panels (b) and (c) show 99th-percentile top whiskers. Panel (d) profiles the event-trace microbenchmark on an Intel Xeon Gold 6140 and annotates IPC.}
  \label{fig:eval-uccl}
\end{figure*}

\begin{figure*}[!t]
  \centering
  \begin{subfigure}[t]{0.24\textwidth}
    \centering
    \includegraphics[width=\linewidth]{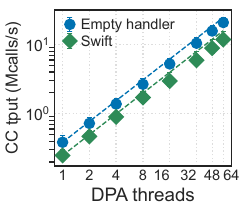}
    \caption{Strong scaling, 3,072 flows.}
    \label{fig:eval-dpa-strong}
  \end{subfigure}\hfill
  \begin{subfigure}[t]{0.24\textwidth}
    \centering
    \includegraphics[width=\linewidth]{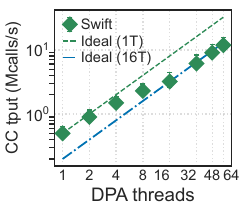}
    \caption{Weak scaling, 48 flows/thread.}
    \label{fig:eval-dpa-weak}
  \end{subfigure}\hfill
  \begin{subfigure}[t]{0.24\textwidth}
    \centering
    \includegraphics[width=\linewidth]{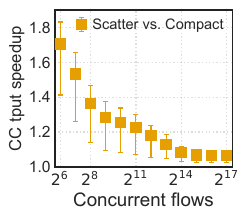}
    \caption{Impact of DPA affinity.}
    \label{fig:eval-dpa-placement}
  \end{subfigure}\hfill
  \begin{subfigure}[t]{0.24\textwidth}
    \centering
    \includegraphics[width=\linewidth]{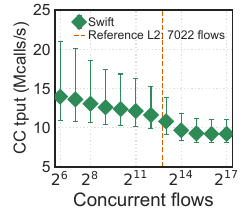}
    \caption{Flow scaling, 64 threads.}
    \label{fig:eval-dpa-compact}
  \end{subfigure}
  \caption{PCM-Swift scaling under DPA-saturating trace replay. Markers show means of means across four traces and three seeds. Whiskers span individual means. Dashed lines in panels (a) and (b) show ideal scaling.}
  \label{fig:eval-dpa-scaling}
  \vspace{-1em}
\end{figure*}

\paragraph{Step 4: End-to-end NIC offloading deployment.}
We now evaluate PCM-Swift offloaded to a ConnectX-8. We run a 3:1 incast from three leaf NICs to one root NIC (Figure~\ref{fig:eval-connectx-swift}). Each sender runs \texttt{ib\_write\_bw} with contiguous 64~KiB RDMA Writes across a single isolated 100~Gbit/s plane of an 800~Gbit/s port. Traffic from all senders crosses two tiers of switches toward the root NIC, with an RTT of $\approx15~\mu$s.

\emph{Control instrumentation.} To expose every live PCM-Swift control decision, we enable diagnostic mode, which reports handler outputs after every invocation. Emitting and flushing these records on the DPA callback path adds multiple DPA--host PCIe RTTs to the handler time. We exploit this delay to exercise the event-coalescing mechanism and validate the CWND-to-rate adapter used in our Swift implementation (Section~\ref{sec:cx8-impl}). The next DOCA PCC callback delivers the ACK feedback coalesced during logging to the PCM shell's SUM accumulator.

\emph{Fairness under coarse feedback.} Figure~\ref{fig:eval-connectx-coalescing} shows that the median batch size decreases as senders throttle to their fair shares. A lone sender produces median batches of 110--129 ACKs. With two senders, the median batch contains 63 ACKs. With three senders, it contains 42 ACKs. Despite the coarse instrumented schedule, Figure~\ref{fig:eval-connectx-fairness} shows that PCM-Swift converges to stable fair-share operation as contenders enter and leave. We exclude the first and last second of each phase when averaging pacer rate fairness. In each of the five phases, mean Jain's fairness indices for NIC pacer rate governed by the policy and Write goodput are at least 0.99.  

The production shell build removes per-invocation diagnostics. With one flow per 100~Gbit/s plane, the optimized PCM-Swift handler sustains one invocation per four ACKs. Production reaction intervals also depend on ConnectX's RDMA engine ACK pacing, DOCA PCC's internal callback scheduling and coalescing, and system load. We study the effect of system load in Section~\ref{sec:eval-scaling}. This live 3:1 incast experiment complements the controlled scheduling-delay sweep in simulation by testing Swift under ConnectX's native event coalescing and rate enforcement.

\paragraph{Step 5: Crossing the transport boundary.}

We tested PCM-Swift on Microsoft 1P under end-to-end and synthetic traffic. We observe that it reacts to synthetically injected high RTT samples by reducing its CWND, matching Swift's control response in UET-htsim. We discuss UCCL in detail as an architectural counterpart to NIC offloading on CX-8 and M-1P. Our UCCL study tests whether co-locating PCM preserves line-rate goodput and fine-grained reactions under CPU saturation. Appendix~\ref{sec:uccl-appendix} evaluates PCM-Swift's CWND response to injected RTT spikes.

\emph{Goodput scaling.} We deploy the portable PCM handler in UCCL to control the injection of 64~KiB UC Write chunks, as Figure~\ref{fig:eval-uccl-throughput} shows. We adjust only Swift parameters in the specification to match UCCL's parameters. The handler retains shared policy arithmetic and object interfaces. Built-in Swift and PCM-Swift reach practical RoCEv2 line rate across sixteen planes with four engine workers.

\emph{Handler-call scaling.} Figure~\ref{fig:eval-uccl-placement} shows how the PCM worker's handler-call rate in response to UCCL ACKs changes with the number of dataplane workers. Synchronous PCM tracks the built-in handler-call baseline through four datapath engines and plateaus when the system saturates all sixteen planes. Figure~\ref{fig:eval-uccl-coalescing} shows that the median fraction of invocations with multiple ACK updates is below 5\% on SMT, 13.3\% on a same-NUMA core, and 34.8\% on remote NUMA.

\emph{Control-execution cost.} Figure~\ref{fig:eval-uccl-profile} shows our PCM shell microbenchmark on the Xeon platform. The breakdown shows 66~ns for synchronous execution, 87~ns on SMT, and 192~ns on a separate core. The \texttt{std::chrono} operation accounts for a large part of the snapshot cost and takes $\approx20$~ns. Swift remains near 35--40~ns. The synchronous and asynchronous shells execute nearly the same number of instructions. The cache-coherence protocol of the atomic interconnect decreases IPC from 0.98 (Sync) to 0.77 (Async SMT) and 0.36 (Async same NUMA).

\emph{Cost of alternative designs.} An alternative to PCM's asynchronous control plane is to integrate UCCL with CCP~\cite{narayan2018restructuring}, whose datapath DSL accumulates measurements and triggers reports to host-side policies through IPC. A minimal CCP filter for ACK accumulation and triggering alone compiles to four DSL bytecodes, but glue code executes approximately 880 x86-64 instructions to enter and evaluate the filter, before report generation or IPC. Our separate NewReno profile on the same Xeon measures 364~ns for DSL evaluation. 

In comparison, with operands in registers, PCM-Swift's ACK-count accumulation (SUM) and RTT-sample update (LAST) each compile to one instruction, while its ACK trigger (\texttt{DELTA >= 1}) reduces to a comparison and conditional branch. On the Xeon, signal and control exchange uses single-instruction atomic loads, stores, and additions. Swift shell invocations, including policy execution, take 66--192~ns (Figure~\ref{fig:eval-uccl-profile}). Moving CCP's host-side policy execution onto ConnectX would also require replacing Portus's Rust runtime its Rust policies for the DPA's C11 environment~\cite{doca-pcc}.

\subsection{PCM Offloading Scalability}
\label{sec:eval-scaling}

\begin{figure*}[!t]
  \centering
  \includegraphics[width=\textwidth]{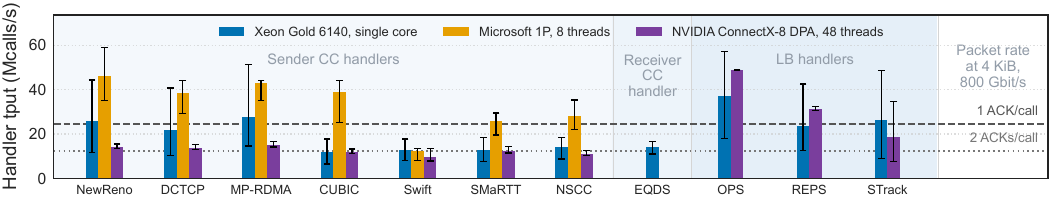}
  \caption{CC and LB handler performance. Whiskers show Microsoft 1P extrema and Intel Xeon 99.9th percentiles. We benchmark Xeon and NVIDIA ConnectX-8 with event traces for 192 flows. 1P runs live RDMA Write traffic across 16 QPs.}
  \label{fig:eval-backends}
\end{figure*}

We analyze PCM offload performance beyond the scale of end-to-end traffic in our four-NIC ConnectX testbed. Figure~\ref{fig:eval-dpa-scaling} reports handler performance.

\emph{Strong scaling.} In Figure~\ref{fig:eval-dpa-strong}, the working-set size remains fixed across threads. The empty handler and Swift reach 75--85\% of ideal throughput. With 64 of the DPA's 256 hardware threads and 3072 flows, PCM-Swift sustains 11.83 \mbox{Mcalls/s}.

\emph{Weak scaling and thread affinity.} Starting at 16 threads, PCM-Swift achieves 30--40\% of the ideal throughput extrapolated from the single-thread reference (1T) in Figure~\ref{fig:eval-dpa-weak}. Performance then stabilizes. At 64 threads, Swift uses four fully subscribed DPA cores. Swift reaches 90\% of a weak-scaling curve derived from the 16-thread (16T) reference. This reference saturates all DPA core threads.

This result signifies the effect of the DPA's SMT microarchitecture. Under the default compact mapping, increasing the thread count from 1 to 16 gradually increases pressure on the same ALU. Figure~\ref{fig:eval-dpa-placement} reports the relative speedup of scattered mappings over compact mappings and shows the impact of shared-ALU pressure. With 4,096 flows, 64 scattered threads, and four threads per ALU, PCM-Swift reaches 13.5 \mbox{Mcalls/s}. Compact mapping reaches 11.6 \mbox{Mcalls/s}. Our Microsoft 1P experiments, which assign four hardware threads to one ALU, exhibit the same trend.

\emph{Flow scaling.} Figure~\ref{fig:eval-dpa-compact} shows Swift throughput decreasing from 11.61 to 10.81 \mbox{Mcalls/s} as the working set grows from 0.875 to 1.75~MiB. To interpret this trend, we use published BlueField-3 DPA memory-hierarchy measurements as a reference~\cite{chen2024demystifying}. Its reported L2 capacity corresponds to $>$7000 Swift flow contexts, marked by the vertical guide. The observed decline is consistent with increased cache pressure near this reference capacity. At 131072 flows, compact and scattered placements both sustain nearly 9 \mbox{Mcalls/s}.

\subsection{Performance Across Handlers}
\label{sec:handlers-perf}

We evaluate PCM handlers across the supported shells in Figure~\ref{fig:eval-backends}. This experiment tests shell reuse across policies, policy portability across shells, and overall handler-performance trends across platforms and algorithms. For the asynchronous CPU shell, the trace microbenchmark saturates each worker, as does the DPA benchmark.

One Xeon core sustains 11.8--37.3 million asynchronous controller calls/s, and eight Microsoft 1P threads provide 12.2--46.3 million calls/s.
Falcon reports 20 million FAE events/s on one Neoverse-N1 core, sustained across connection counts through state prefetching~\cite[Fig.~22a]{singhvi2025falcon}. PCM-Swift in Figure~\ref{fig:eval-backends} takes on average 77.1 ns/ACK on Xeon, comparable to the 66 ns RUE service interval modeled by Falcon's simulator, which presumably models the performance of Neoverse-N1 cores. This result provides architectural context rather than a matched performance baseline.

ConnectX's 48 DPA threads sustain 9.5--15.1 million CC calls/s. For the simpler, less compute-intensive LB shell, the same 48 DPA threads sustain 18.6--48.8 million LB calls/s. For a lightweight scheme such as MP-RDMA, discussed in Section~\ref{sec:api-examples}, these capacities approach per-packet policy demand on an 800~Gbit/s link. For Swift, the most compute-intensive algorithm, the capacity corresponds to one input event per two to three ACKs.

\subsection{Lessons Learned}

\paragraph{Optimize policies and backends independently.} PCM demonstrates two levels of reuse. Vendors optimize backend arithmetic to accelerate native execution across policies, while developers carry algebraic improvements in portable handlers across supported platforms.

\paragraph{Validate numerical behavior and reaction timing.} Swift’s fixed-point failure and scheduling sensitivity show why source compatibility needs behavioral validation. Accumulation preserves feedback counts, but delayed reactions can still reduce fairness and increase flow completion times. Use simulation to validate numerical representations and establish a workload-specific reaction budget.

\paragraph{Provision NIC compute for the required reaction rate.} Thread placement and flow-state footprint shape processing capacity: hardware threads share execution resources, and growing state increases cache pressure. Provision workers to meet the required reaction rate within these constraints. 

\raggedbottom

\section{Conclusion}

PCM makes portable congestion management practical for high-performance RDMA. Our programming model spans eleven policies and four datapaths, from simulation and software transports to NIC offload. PCM policies sustain per-chunk control at 1.56~Tbit/s goodput in UCCL and achieve near-per-ACK processing capacity on vendor 400G and 800G NICs. With its design informing ongoing Ultra Ethernet standardization, PCM opens a path for congestion-control and load-balancing innovations to reach heterogeneous datacenters through a shared interface.

\section*{Acknowledgments}
We thank Shuhao Li, Zhiyi Hu, Inho Cho, Nadeen Gebara, Luca Piccolboni, and Terry Lam for their feedback and assistance with the experiments. This project received funding from the European Research Council (ERC) under the European Union’s Horizon 2020 research and innovation program (grant agreement PSAP, No. 101002047) and a CAF America grant. We also thank the Swiss National Supercomputing Center (CSCS) for providing the computational resources used in this work.

\bibliographystyle{acm}
\bibliography{literature}

@article{li2020pytorch,
  title={Pytorch distributed: Experiences on accelerating data parallel training},
  author={Li, Shen and Zhao, Yanli and Varma, Rohan and Salpekar, Omkar and Noordhuis, Pieter and Li, Teng and Paszke, Adam and Smith, Jeff and Vaughan, Brian and Damania, Pritam and others},
  journal={arXiv preprint arXiv:2006.15704},
  year={2020}
}

@article{ma2026challenges,
  title={Challenges and research directions for large language model inference hardware},
  author={Ma, Xiaoyu and Patterson, David},
  journal={Computer},
  volume={59},
  number={5},
  pages={55--64},
  year={2026},
  publisher={IEEE}
}

@inproceedings{lei2026fast,
  title={$\{$FAST$\}$: An Efficient Scheduler for $\{$All-to-All$\}$$\{$GPU$\}$ Communication},
  author={Lei, Yiran and Lee, Dongjoo and Zhao, Liangyu and Kurniawan, Daniar and Kim, Chanmyeong and Jeong, Heetaek and Kim, Changsu and Choi, Hyeonseong and Yu, Liangcheng and Krishnamurthy, Arvind and others},
  booktitle={23rd USENIX Symposium on Networked Systems Design and Implementation (NSDI 26)},
  pages={2515--2531},
  year={2026}
}

@inproceedings{jiang2026odin,
  title={Odin: Rethinking Congestion Control under All-to-All Traffic},
  author={Jiang, Wanchun and Hu, Yujie and Wang, Kai and Han, Xiao and Wei, Haoran and Zhang, Yong and Ye, Jianxi and Dai, Huichen},
  booktitle={Proceedings of the ACM SIGCOMM 2026 Conference},
  pages={1246--1263},
  year={2026}
}

@inproceedings{shashidhara2026presto,
  title={Presto: A Match-Action TCP Stack for the Terabit Era},
  author={Shashidhara, Rajath and Kaufmann, Antoine and Peter, Simon},
  booktitle={Proceedings of the ACM SIGCOMM 2026 Conference},
  pages={1360--1375},
  year={2026}
}

@article{zhao2023pytorch,
  title={Pytorch fsdp: experiences on scaling fully sharded data parallel},
  author={Zhao, Yanli and Gu, Andrew and Varma, Rohan and Luo, Liang and Huang, Chien-Chin and Xu, Min and Wright, Less and Shojanazeri, Hamid and Ott, Myle and Shleifer, Sam and others},
  journal={arXiv preprint arXiv:2304.11277},
  year={2023}
}

@article{hoefler2025ultra,
  title={Ultra Ethernet's Design Principles and Architectural Innovations},
  author={Hoefler, Torsten and Schramm, Karen and Spada, Eric and Underwood, Keith and Alexander, Cedell and Alverson, Bob and Bottorff, Paul and Caulfield, Adrian and Handley, Mark and Huang, Cathy and others},
  journal={arXiv preprint arXiv:2508.08906},
  year={2025}
}

@article{ha2008cubic,
  title={CUBIC: a new TCP-friendly high-speed TCP variant},
  author={Ha, Sangtae and Rhee, Injong and Xu, Lisong},
  journal={ACM SIGOPS operating systems review},
  volume={42},
  number={5},
  pages={64--74},
  year={2008},
  publisher={ACM New York, NY, USA}
}

@inproceedings{huang2025swcc,
  title={$\{$SwCC$\}$:$\{$Software-Programmable$\}$ and $\{$Per-Packet$\}$ Congestion Control in $\{$RDMA$\}$ Engine},
  author={Huang, Hongjing and Zhang, Jie and Chen, Xuzheng and Song, Ziyu and Qin, Jiajun and Wang, Zeke},
  booktitle={2025 USENIX Annual Technical Conference (USENIX ATC 25)},
  pages={1243--1260},
  year={2025}
}

@inproceedings{olteanu2022edge,
  title={An edge-queued datagram service for all datacenter traffic},
  author={Olteanu, Vladimir and Eran, Haggai and Dumitrescu, Dragos and Popa, Adrian and Baciu, Cristi and Silberstein, Mark and Nikolaidis, Georgios and Handley, Mark and Raiciu, Costin},
  booktitle={19th USENIX Symposium on Networked Systems Design and Implementation (NSDI 22)},
  pages={761--777},
  year={2022}
}

@inproceedings{gangidi2024rdma,
  title={Rdma over ethernet for distributed training at meta scale},
  author={Gangidi, Adithya and Miao, Rui and Zheng, Shengbao and Bondu, Sai Jayesh and Goes, Guilherme and Morsy, Hany and Puri, Rohit and Riftadi, Mohammad and Shetty, Ashmitha Jeevaraj and Yang, Jingyi and others},
  booktitle={Proceedings of the ACM SIGCOMM 2024 Conference},
  pages={57--70},
  year={2024}
}

@inproceedings{zeng2026connecting,
  title={Connecting 100K+ GPUs: Building the Communication Stack for Large-Scale LLM Training},
  author={Zeng, Hongyi and Si, Min and Balaji, Pavan and Chen, Yongzhou and Chu, Ching-Hsiang and Gangidi, Adithya and Kannan, Prashanth and Liu, Bingzhe and Hasan, Saif and He, Dong and others},
  booktitle={Proceedings of the ACM SIGCOMM 2026 Conference},
  pages={505--518},
  year={2026}
}

@rfc{ecmp,
author = {Hopps, C.},
title = {RFC2992: Analysis of an Equal-Cost Multi-Path Algorithm},
year = {2000},
publisher = {RFC Editor},
address = {USA}
}

@ARTICLE{sdr,
  author={Shalev, Leah and Ayoub, Hani and Bshara, Nafea and Sabbag, Erez},
  journal={IEEE Micro}, 
  title={A Cloud-Optimized Transport Protocol for Elastic and Scalable HPC}, 
  year={2020},
  volume={40},
  number={6},
  pages={67-73},
  doi={10.1109/MM.2020.3016891}}

@inproceedings{zhao2025white,
  title={$\{$White-Boxing$\}$$\{$RDMA$\}$ with $\{$Packet-Granular$\}$ Software Control},
  author={Zhao, Chenxingyu and Min, Jaehong and Liu, Ming and Krishnamurthy, Arvind},
  booktitle={22nd USENIX Symposium on Networked Systems Design and Implementation (NSDI 25)},
  pages={427--449},
  year={2025}
}

@article{le2024strack,
  title={STrack: A Reliable Multipath Transport for AI/ML Clusters},
  author={Le, Yanfang and Pan, Rong and Newman, Peter and Blendin, Jeremias and Kabbani, Abdul and Jain, Vipin and Sivaramu, Raghava and Matus, Francis},
  journal={arXiv preprint arXiv:2407.15266},
  year={2024}
}

@inproceedings{uccl_transport,
  title={$\{$UCCL-Tran$\}$: An Extensible Software Transport Layer for $\{$GPU$\}$ Networking},
  author={Zhou, Yang and Chen, Zhongjie and Mao, Ziming and Lao, ChonLam and Yang, Shuo and Kannan, Pravein Govindan and Zhang, Xizhi and Gao, Jiaqi and Zhao, Yilong and Wu, Yongji and others},
  booktitle={20th USENIX Symposium on Operating Systems Design and Implementation (OSDI 26)},
  pages={1143--1166},
  year={2026}
}

@inproceedings{arashloo2020enabling,
  title={Enabling programmable transport protocols in $\{$High-Speed$\}$$\{$NICs$\}$},
  author={Arashloo, Mina Tahmasbi and Lavrov, Alexey and Ghobadi, Manya and Rexford, Jennifer and Walker, David and Wentzlaff, David},
  booktitle={17th USENIX Symposium on Networked Systems Design and Implementation (NSDI 20)},
  pages={93--109},
  year={2020}
}

@inproceedings{arashloo2017hotcocoa,
  title={Hotcocoa: Hardware congestion control abstractions},
  author={Arashloo, Mina Tahmasbi and Ghobadi, Monia and Rexford, Jennifer and Walker, David},
  booktitle={Proceedings of the 16th ACM Workshop on Hot Topics in Networks},
  pages={108--114},
  year={2017}
}

@article{dean2013tail,
  title={The tail at scale},
  author={Dean, Jeffrey and Barroso, Luiz Andr{\'e}},
  journal={Communications of the ACM},
  volume={56},
  number={2},
  pages={74--80},
  year={2013},
  publisher={ACM New York, NY, USA}
}

@inproceedings{dcqcn,
author = {Zhu, Yibo and Eran, Haggai and Firestone, Daniel and Guo, Chuanxiong and Lipshteyn, Marina and Liron, Yehonatan and Padhye, Jitendra and Raindel, Shachar and Yahia, Mohamad Haj and Zhang, Ming},
title = {Congestion Control for Large-Scale RDMA Deployments},
year = {2015},
isbn = {9781450335423},
publisher = {Association for Computing Machinery},
address = {New York, NY, USA},
url = {https://doi.org/10.1145/2785956.2787484},
doi = {10.1145/2785956.2787484},
booktitle = {Proceedings of the 2015 ACM Conference on Special Interest Group on Data Communication},
pages = {523–536},
numpages = {14},
location = {London, United Kingdom},
series = {SIGCOMM '15}
}

@inproceedings{spark,
author = {Zaharia, Matei and Chowdhury, Mosharaf and Das, Tathagata and Dave, Ankur and Ma, Justin and McCauley, Murphy and Franklin, Michael J. and Shenker, Scott and Stoica, Ion},
title = {Resilient distributed datasets: a fault-tolerant abstraction for in-memory cluster computing},
year = {2012},
publisher = {USENIX Association},
address = {USA},
booktitle = {Proceedings of the 9th USENIX Conference on Networked Systems Design and Implementation},
pages = {2},
numpages = {1},
location = {San Jose, CA},
series = {NSDI'12}
}

@article{mapreduce,
author = {Dean, Jeffrey and Ghemawat, Sanjay},
title = {MapReduce: simplified data processing on large clusters},
year = {2008},
issue_date = {January 2008},
publisher = {Association for Computing Machinery},
address = {New York, NY, USA},
volume = {51},
number = {1},
issn = {0001-0782},
url = {https://doi.org/10.1145/1327452.1327492},
doi = {10.1145/1327452.1327492},
journal = {Commun. ACM},
month = jan,
pages = {107–113},
numpages = {7}
}

@inproceedings {memcache,
author = {Rajesh Nishtala and Hans Fugal and Steven Grimm and Marc Kwiatkowski and Herman Lee and Harry C. Li and Ryan McElroy and Mike Paleczny and Daniel Peek and Paul Saab and David Stafford and Tony Tung and Venkateshwaran Venkataramani},
title = {Scaling Memcache at Facebook},
booktitle = {10th USENIX Symposium on Networked Systems Design and Implementation (NSDI 13)},
year = {2013},
isbn = {978-1-931971-00-3},
address = {Lombard, IL},
pages = {385--398},
url = {https://www.usenix.org/conference/nsdi13/technical-sessions/presentation/nishtala},
publisher = {USENIX Association},
month = apr
}

@inproceedings{uno,
author = {Bonato, Tommaso and Abdous, Sepehr and Kabbani, Abdul and Ghalayini, Ahmad and Gebara, Nadeen and Lam, Terry and Agarwal, Anup and Chen, Tiancheng and Yu, Zhuolong and Taranov, Konstantin and Elhaddad, Mahmoud and De Sensi, Daniele and Ghorbani, Soudeh and Hoefler, Torsten},
title = {Uno: A One-Stop Solution for Inter- and Intra-Data Center Congestion Control and Reliable Connectivity},
year = {2025},
isbn = {9798400714665},
publisher = {Association for Computing Machinery},
address = {New York, NY, USA},
url = {https://doi.org/10.1145/3712285.3759884},
doi = {10.1145/3712285.3759884},
booktitle = {Proceedings of the International Conference for High Performance Computing, Networking, Storage and Analysis},
pages = {1195–1210},
numpages = {16},
location = {
},
series = {SC '25}
}

@inproceedings{de2020depth,
  title={An in-depth analysis of the slingshot interconnect},
  author={De Sensi, Daniele and Di Girolamo, Salvatore and McMahon, Kim H and Roweth, Duncan and Hoefler, Torsten},
  booktitle={SC20: International Conference for High Performance Computing, Networking, Storage and Analysis},
  pages={1--14},
  year={2020},
  organization={IEEE}
}

@inproceedings{bonato2025repsrecycledentropypacket,
  title={Reps: Recycled entropy packet spraying for adaptive load balancing and failure mitigation},
  author={Bonato, Tommaso and Kabbani, Abdul and Ghalayini, Ahmad and Papamichael, Michael and Dohadwala, Mohammad and Gianinazzi, Lukas and Khalilov, Mikhail and Achermann, Elias and De Sensi, Daniele and Hoefler, Torsten},
  booktitle={Proceedings of the 21st European Conference on Computer Systems},
  pages={225--246},
  year={2026}
}

@inproceedings{ndp,
author = {Handley, Mark and Raiciu, Costin and Agache, Alexandru and Voinescu, Andrei and Moore, Andrew W. and Antichi, Gianni and W\'{o}jcik, Marcin},
title = {Re-architecting datacenter networks and stacks for low latency and high performance},
year = {2017},
isbn = {9781450346535},
publisher = {Association for Computing Machinery},
address = {New York, NY, USA},
url = {https://doi.org/10.1145/3098822.3098825},
doi = {10.1145/3098822.3098825},
booktitle = {Proceedings of the Conference of the ACM Special Interest Group on Data Communication},
pages = {29–42},
numpages = {14},
location = {Los Angeles, CA, USA},
series = {SIGCOMM '17}
}

@inproceedings {mprdma,
author = {Yuanwei Lu and Guo Chen and Bojie Li and Kun Tan and Yongqiang Xiong and Peng Cheng and Jiansong Zhang and Enhong Chen and Thomas Moscibroda},
title = {{Multi-Path} Transport for {RDMA} in Datacenters},
booktitle = {15th USENIX Symposium on Networked Systems Design and Implementation (NSDI 18)},
year = {2018},
isbn = {978-1-939133-01-4},
address = {Renton, WA},
pages = {357--371},
url = {https://www.usenix.org/conference/nsdi18/presentation/lu},
publisher = {USENIX Association},
month = apr
}

@ARTICLE{geminiTOn,
author={Zeng, Gaoxiong and Bai, Wei and Chen, Ge and Chen, Kai and Han, Dongsu and Zhu, Yibo and Cui, Lei},
journal={ IEEE/ACM Transactions on Networking },
title={{ Congestion Control for Cross-Datacenter Networks }},
year={2022},
volume={30},
number={05},
ISSN={1558-2566},
pages={2074-2089},
doi={10.1109/TNET.2022.3161580},
url = {https://doi.ieeecomputersociety.org/10.1109/TNET.2022.3161580},
publisher={IEEE Computer Society},
address={Los Alamitos, CA, USA},
month=oct}

@article{bbr,
author = {Cardwell, Neal and Cheng, Yuchung and Gunn, C. Stephen and Yeganeh, Soheil Hassas and Van Jacobson},
title = {BBR: congestion-based congestion control},
year = {2017},
issue_date = {February 2017},
publisher = {Association for Computing Machinery},
address = {New York, NY, USA},
volume = {60},
number = {2},
issn = {0001-0782},
url = {https://doi.org/10.1145/3009824},
doi = {10.1145/3009824},
journal = {Commun. ACM},
month = jan,
pages = {58–66},
numpages = {9}
}

@inproceedings{dctcp,
author = {Alizadeh, Mohammad and Greenberg, Albert and Maltz, David A. and Padhye, Jitendra and Patel, Parveen and Prabhakar, Balaji and Sengupta, Sudipta and Sridharan, Murari},
title = {Data center TCP (DCTCP)},
year = {2010},
isbn = {9781450302012},
publisher = {Association for Computing Machinery},
address = {New York, NY, USA},
url = {https://doi.org/10.1145/1851182.1851192},
doi = {10.1145/1851182.1851192},
booktitle = {Proceedings of the ACM SIGCOMM 2010 Conference},
pages = {63–74},
numpages = {12},
location = {New Delhi, India},
series = {SIGCOMM '10}
}

@inproceedings{narayan2018restructuring,
  title={Restructuring endpoint congestion control},
  author={Narayan, Akshay and Cangialosi, Frank and Raghavan, Deepti and Goyal, Prateesh and Narayana, Srinivas and Mittal, Radhika and Alizadeh, Mohammad and Balakrishnan, Hari},
  booktitle={Proceedings of the 2018 Conference of the ACM Special Interest Group on Data Communication},
  pages={30--43},
  year={2018}
}

@inproceedings{swift,
  title={Swift: Delay is simple and effective for congestion control in the datacenter},
  author={Kumar, Gautam and Dukkipati, Nandita and Jang, Keon and Wassel, Hassan MG and Wu, Xian and Montazeri, Behnam and Wang, Yaogong and Springborn, Kevin and Alfeld, Christopher and Ryan, Michael and others},
  booktitle={Proceedings of the Annual conference of the ACM Special Interest Group on Data Communication on the applications, technologies, architectures, and protocols for computer communication},
  pages={514--528},
  year={2020}
}

@inproceedings{singhvi2025falcon,
  title={Falcon: A reliable, low latency hardware transport},
  author={Singhvi, Arjun and Dukkipati, Nandita and Chandra, Prashant and Wassel, Hassan MG and Sharma, Naveen Kr and Rebello, Anthony and Schuh, Henry and Kumar, Praveen and Montazeri, Behnam and Bansod, Neelesh and others},
  booktitle={Proceedings of the ACM SIGCOMM 2025 Conference},
  pages={248--263},
  year={2025}
}

@misc{broadcom-nic,
  title = {P1800GO - 1 x 800GbE PCIe NIC},
  author = {Broadcom},
  year = {2025},
  howpublished = {\url{https://www.broadcom.com/products/ethernet-connectivity/network-adapters/p1800go}},
}

@misc{amd-nic,
  title = {Pensando Vulcano 800 AI NIC},
  author = {AMD},
  year = {2026},
  howpublished = {\url{https://www.amd.com/content/dam/amd/en/documents/pensando-business-docs/product-brief/amd-vulcano-ai-nic-networking-product-brief.pdf}},
}

@misc{ue-htsim,
  title = {uec-transport-simulation-code},
  author = {Ultra Ethernet Consortium},
  year = {2025},
  howpublished = {\url{https://github.com/ultraethernet/uet-htsim}},
}

@inproceedings{chen2024demystifying,
  title={Demystifying datapath accelerator enhanced off-path smartnic},
  author={Chen, Xuzheng and Zhang, Jie and Fu, Ting and Shen, Yifan and Ma, Shu and Qian, Kun and Zhu, Lingjun and Shi, Chao and Zhang, Yin and Liu, Ming and others},
  booktitle={2024 IEEE 32nd International Conference on Network Protocols (ICNP)},
  pages={1--12},
  year={2024},
  organization={IEEE}
}

@misc{ietf-newreno,
  title = {RFC 6582},
  author = {IETF},
  howpublished = {\url{https://datatracker.ietf.org/doc/rfc6582/}},
}

@misc{doca-pcc,
  title = {DOCA PCC},
  author = {NVIDIA},
  howpublished = {\url{https://docs.nvidia.com/doca/sdk/doca-pcc/index.html}},
}

@misc{uec-spec,
  title = {Ultra Ethernet Specification v1.0.1},
  author = {Ultra Ethernet Consortium},
  howpublished = {\url{https://ultraethernet.org/wp-content/uploads/sites/20/2025/10/UE-Specification-1.0.1.pdf}},
}

@inproceedings{hinz2023tcp,
  title={TCP's Third Eye: Leveraging eBPF for Telemetry-Powered Congestion Control},
  author={Hinz, J{\"o}rn-Thorben and Addanki, Vamsi and Gy{\"o}rgyi, Csaba and Jepsen, Theo and Schmid, Stefan},
  booktitle={Proceedings of the 1st Workshop on eBPF and Kernel Extensions},
  pages={1--7},
  year={2023}
}

@article{smartt,
  title={SMaRTT: Sender-based Marked Rapidly-adapting Trimmed \& Timed Transport},
  author={Bonato, Tommaso and Kabbani, Abdul and Ghalayini, Ahmad and Agarwal, Anup and De Sensi, Daniele and Pan, Rong and Raiciu, Costin and Handley, Mark and Brodschi, Mihai and Schneider, Timo and others},
  journal={arXiv preprint arXiv:2404.01630},
  year={2024}
}

@INPROCEEDINGS{rps,
  author={Dixit, Advait and Prakash, Pawan and Hu, Y. Charlie and Kompella, Ramana Rao},
  booktitle={2013 Proceedings IEEE INFOCOM}, 
  title={On the impact of packet spraying in data center networks}, 
  year={2013},
  volume={},
  number={},
  pages={2130-2138},
  doi={10.1109/INFCOM.2013.6567015}}

@inproceedings{lattner2004llvm,
  title={LLVM: A compilation framework for lifelong program analysis \& transformation},
  author={Lattner, Chris and Adve, Vikram},
  booktitle={International symposium on code generation and optimization, 2004. CGO 2004.},
  pages={75--86},
  year={2004},
  organization={IEEE}
}

\clearpage
\appendix
\section{Behavioral Validation of PCM-Swift in UCCL}
\label{sec:uccl-appendix}

We compare the CWND behavior of PCM-Swift, deployed through the reference CPU shell, with that of native UCCL Swift.
We evaluate both synchronous and asynchronous PCM integrations with UCCL.
The asynchronous CPU shell supports relaxed and locked snapshot modes (Section~\ref{sec:exec-modes}).
In locked snapshot mode, the datapath brackets atomic signal updates with sequence numbers.
The control plane worker retries its sequential signal reads until the same even sequence number precedes and follows the copy.
Relaxed snapshot mode accepts the first copy.

\begin{figure}[htbp]
  \centering
  \includegraphics[width=\linewidth]{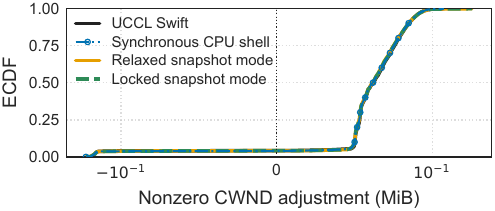}
  \caption{Nonzero CWND adjustments under RTT injection. Each curve uses 400{,}000 datapath observations from a single run.}
  \label{fig:appendix-skew-effect}
\end{figure}

We model transient network congestion by injecting a synthetic $500~\mu$s RTT every 25th CC update.
Figure~\ref{fig:appendix-skew-effect} compares nonzero window adjustments under RTT injection for native UCCL Swift, the synchronous CPU shell, and both asynchronous snapshot modes.
In the asynchronous runs, the control plane worker runs on a separate core.
Across all modes, the range of PCM-Swift's CWND adjustments closely tracks that of native UCCL Swift.

The median nonzero adjustments in relaxed and locked snapshot modes differ by 1.2\%.
Their 1st and 99th percentiles differ by less than 1\%.
The datapath observes an unchanged window in 10.9\% of samples in relaxed snapshot mode and 10.7\% in locked snapshot mode.
When the worker runs on the datapath thread's SMT sibling, median nonzero adjustments in relaxed and locked snapshot modes differ by 0.7\%.

\newpage

\section{Specification for the MP-RDMA Handler}
\label{sec:spec-appendix}
\begin{lstlisting}[style=pcmcode,language=PCMYAML,float=htbp,caption={ACK-handling specification for Listing~\ref{lst:cc}.},label={lst:cc-yaml}]
controls:
  - name: CWND
    control: CWND_BYTES
signals:
  - name: ACKED_BYTES
    type: DATA_ACK_BYTES
    accumulation: SUM
    trigger_type: DELTA
    trigger_op: >=
    trigger_threshold: 1
  - name: NACKED_BYTES
    type: DATA_NACK_BYTES
    accumulation: SUM
    trigger_type: DELTA
    trigger_op: >=
    trigger_threshold: 1
variables:
  - name: PREV_ACKED_BYTES
    type: uint
    initial_value: 0
  - name: PREV_NACKED_BYTES
    type: uint
    initial_value: 0
constants:
  - name: MTU
    type: uint
    value: 4096
\end{lstlisting}

\end{document}